\documentclass[twocolumn, journal]{IEEEtran}

\IEEEoverridecommandlockouts

\usepackage{color}
\usepackage{graphicx}
\usepackage{amsmath}
\usepackage{amssymb}
\usepackage{algorithm}
\usepackage{algorithmic}

\newcommand{\be}{\begin{equation}}
\newcommand{\ee}{\end{equation}}
\newcommand{\bea}{\begin{eqnarray}}
\newcommand{\eea}{\end{eqnarray}}
\newcommand{\ba}{\begin{array}}
\newcommand{\ea}{\end{array}}
\newcommand{\nid}{\noindent}
\newcommand{\non}{\nonumber}

\renewcommand{\thesubsubsection}{\thesubsection.\arabic{subsubsection}}

\title{Hybrid Precoder and Combiner Design with Low Resolution Phase Shifters in mmWave MIMO Systems}

\author{Zihuan Wang,~\IEEEmembership{Student Member,~IEEE,}
          Ming Li,~\IEEEmembership{Senior Member,~IEEE,}
        Qian Liu,~\IEEEmembership{Member,~IEEE,}
         and A. Lee Swindlehurst,~\IEEEmembership{Fellow,~IEEE}
         \thanks{Zihuan Wang and Ming Li are with the School of Information and Communication Engineering, Dalian University of Technology, Dalian, Liaoning 116024, China, (e-mail: wangzihuan@mail.dlut.edu.cn, mli@dlut.edu.cn).}
\thanks{Qian Liu is with the School of Computer Science and Technology, Dalian University of Technology, Dalian, Liaoning 116024, China (e-mail: qianliu@dlut.edu.cn).}
\thanks{A. L. Swindlehurst is with the Center for Pervasive Communications and Computing, University of California, Irvine, CA 92697 USA and also with the Institute for Advanced Study, Technical University of Munich,Munchen 80333, Germany (e-mail: swindle@uci.edu).}

\thanks{This paper is supported by the National Natural Science Foundation of China (Grant No. 61671101 and 61601080) and the Fundamental Research Funds for the Central Universities (Grant No. DUT15RC(3)121 and DUT17JC10).}
}

\begin{document}

\pagestyle{empty}

\maketitle

\begin{abstract}
Millimeter wave (mmWave) communications have been considered as a key technology for next generation cellular systems and Wi-Fi networks because of its advances in providing orders-of-magnitude wider bandwidth than current wireless networks.
Economical and energy-efficient analog/digial hybrid precoding and combining transceivers have been often proposed for mmWave massive multiple-input multiple-output (MIMO) systems to overcome the severe propagation loss of mmWave channels.
One major shortcoming of existing solutions lies in the assumption of infinite or high-resolution phase shifters (PSs) to realize the analog beamformers.
However, low-resolution PSs are typically adopted in practice to reduce the hardware cost and power consumption. Motivated by this fact, in this paper, we investigate the practical design of hybrid precoders and combiners with \textit{low-resolution} PSs in mmWave MIMO systems.
In particular, we propose an iterative algorithm which successively designs the low-resolution analog precoder and combiner pair for each
data stream, aiming at conditionally maximizing the
spectral efficiency.
Then, the digital precoder and combiner are computed based on the obtained effective baseband channel to further enhance the spectral efficiency.
In an effort to achieve an even more hardware-efficient large antenna array, we also investigate the design of hybrid beamformers with \textit{one-bit} resolution (binary) PSs, and present a novel binary analog precoder and combiner optimization algorithm with quadratic complexity in the number of antennas. The proposed low-resolution hybrid beamforming design is further extended to multiuser MIMO communication systems.
Simulation results demonstrate the performance advantages of the proposed algorithms compared to existing low-resolution hybrid beamforming designs, particularly for the one-bit resolution PS scenario.
\end{abstract}


\maketitle

\section{Introduction}

The past decade has witnessed the exponential growth of data traffic along with the rapid proliferation of wireless devices. This flood of mobile traffic has significantly exacerbated spectrum congestion in current frequency bands, and therefore stimulated intensive interest in exploiting new spectrum bands for wireless communications. Millimeter wave (mmWave) wireless communications, operating in the frequency bands from 30-300 GHz, have been demonstrated as a promising candidate to fundamentally solve the spectrum congestion problem \cite{Pi CM 11}-\cite{Swindlehurst 14}.

However, challenges always come along with opportunities. MmWave communications still need to overcome several technical difficulties before real-world deployment. As a negative result of the ten-fold increase of the carrier frequency, the propagation loss in mmWave bands is much higher than that of conventional frequency bands (e.g. 2.4 GHz) due to atmospheric absorption, rain attenuation, and low penetration \cite{Rappaprot 15}.
From a positive perspective, the smaller wavelength of mmWave signals allows a large antenna array to be packed in a small physical dimension \cite{Heath 16}. With the aid of pre/post-coding techniques in massive multiple-input multiple-output (MIMO) systems, the large antenna array can provide sufficient beamforming gain to overcome the severe propagation loss of mmWave channels. It also enables simultaneous transmission of multiple data streams resulting in significant improvements to spectral efficiency.

For MIMO systems operating in conventional cellular frequency bands, the full-digital precoder and combiner are completely realized
in the digital domain by adjusting both the magnitude and phase of the baseband signals. However, these conventional full-digital schemes require a large number of expensive and energy-intensive radio frequency (RF) chains, analog-to-digital converters (ADCs), and digital-to-analog converters (DACs). Since mmWave communication systems operate at much higher carrier frequencies and wider bandwidths, the enormous cost and power consumption of the required RF chains and ADCs/DACs make the adoption of full-digital precoding and combining schemes impractical for mmWave systems.
Recently, economical and energy-efficient analog/digital hybrid precoders and combiners have been advocated as a promising approach to tackle this issue.
The hybrid precoding approaches adopt a large number of phase shifters (PSs) to implement high-dimensional analog precoders to compensate for the severe path-loss at mmWave bands, and a small number of RF chains and DACs to realize low-dimensional digital precoders to provide the necessary flexibility to perform advanced multiplexing/multiuser techniques.

The investigation of hybrid precoder and combiner design has attracted extensive attention in recent years because of its potential energy efficiency for mmWave MIMO communications. The major challenges in designing hybrid precoders are the practical constraints associated with the analog components, such as the requirement that the analog precoding be implemented with constant modulus PSs.
Thus, hybrid precoder design typically requires the solution of various matrix factorization problems with constant modulus constraints.
In particular, a popular solution to maximize the spectral efficiency of point-to-point transmission is to minimize the Euclidean distance between the hybrid precoder and the full-digital precoder \cite{Yu JSAC 16}-\cite{Ni TSP 17}.
Hybrid precoder design for partially-connected architectures are also studied in \cite{Gao JSAC 16}-\cite{Shiwen Access 16}.
Due to the special characteristics of mmWave channels, codebook-based hybrid precoder designs are commonly proposed \cite{Ayach TWC 14}-\cite{Gao TVT 16}, in which the columns of the analog precoder are selected from certain candidate vectors, such as array response
vectors of the channel and discrete Fourier transform (DFT) beamformers. Extensions of the hybrid beamformer design to multiuser mmWave MIMO systems have also been investigated in \cite{Han CM 15}-\cite{Bogale TWC 16}.

The aforementioned existing hybrid precoder and combiner designs generally assume that infinite or high-resolution PSs are used for implementing the analog beamformers in order to achieve satisfactory performance close to the full-digital scheme.
However, implementing infinite/high-resolution PSs at mmWave frequencies would significantly increase the energy consumption and complexity of the required hardware circuits \cite{Poon 12}, \cite{Rial Access}.
Obviously, it is impractical to employ infinite/high-resolution PSs for mmWave systems and real-world analog beamformers will be implemented with low-resolution PSs. Consequently, an important research direction is the exploration of signal processing techniques for hybrid analog/digital architectures that can mitigate the loss of beamforming accuracy due to the low-resolution PSs.

A straightforward approach to obtain the finite-resolution beamformer is to design the infinite-resolution analog beamformer first, and then directly quantize each phase term to a finite set \cite{Chen TVT 17}.
However, this solution becomes inefficient when the PSs have very low resolution.
An alternative solution for hybrid beamforming with finite-resolution PSs is codebook-based design \cite{Ayach TWC 14}-\cite{Gao TVT 16}.
However, for low-resolution PSs, the size of the codebook is very small and the resulting performance is not satisfactory.
In \cite{Sohrabi 15}, \cite{Sohrabi 16}, Sohrabi and Yu proposed to iteratively design the low-resolution hybrid precoder to maximize the spectral efficiency. However, the performance of this algorithm often suffers when one-bit quantized PSs are applied.

In this paper, we first consider the problem of designing hybrid precoders and combiners with low-resolution PSs for a point-to-point mmWave MIMO system.
The objective of the proposed algorithm is to minimize the performance loss caused by the low-resolution PSs while maintaining a low computational complexity.
To achieve this goal, we propose to successively design the low-resolution analog precoder and combiner pair for each data stream, aiming at conditionally maximizing the spectral efficiency.
An iterative phase matching algorithm is introduced to implement the low-resolution analog precoder and combiner pair.
Then, the digital precoder and combiner are computed based on the obtained effective baseband channel to further
enhance the spectral efficiency.

Note that the power consumption and cost of the PS are proportional to its resolution. For example, a 4-bit (i.e. $22.5^{\circ}$) resolution PS at mmWave frequencies requires 45-106 mW, while a 3-bit (i.e. $45^{\circ}$) resolution PS needs only 15 mW  \cite{Rial Access}.
In an effort to achieve maximum hardware efficiency, we also investigate the design of hybrid beamformers with one-bit resolution (binary) PSs.
Inspired by the findings in \cite{Karystinos 07}, we present a binary analog precoder and combiner optimization algorithm under a rank-1 approximation of the interference-included equivalent channel.
This algorithm has quadratic complexity in the number of antennas and can achieve almost the same performance as the optimal exhaustive search method.
Finally, our investigation of low-resolution hybrid precoders and combiners is extended to multiuser mmWave MIMO systems.
Numerical results in the simulation section demonstrate that the proposed algorithms can offer a performance improvement compared with existing low-resolution hybrid beamforming schemes, especially for the one-bit resolution PS scenario.

%
%

\textit{Notation:} The following notation is used throughout this paper. Boldface
lower-case and upper-case letters indicate column vectors and matrices, respectively. $(\cdot)^T$ and $(\cdot)^H$  denote  the transpose and
transpose-conjugate operations, respectively.
$\mathbb{E} \{ \cdot \}$ represents
statistical expectation. $\mathfrak{Re} \{ \cdot \}$ extracts the real part of a complex number; $\mathrm{sign}( \cdot )$ denotes the sign operator; $\mathrm{angle} \{ \cdot \}$ represents the phase of a complex number.
 $\mathbf{I}_L$ indicates an $L \times L$ identity matrix. $\mathbb{C}$ denotes the set of complex numbers.
 $| \mathbf{A} |$ denotes the determinant of matrix $\mathbf{A}$. $| \mathcal{A} |$  denotes the cardinality of set $\mathcal{A}$.  $| a |$ and $\| \mathbf{a} \|$ are the magnitude and norm of a scalar $a$ and vector $\mathbf{a}$, respectively.  $\| \mathbf{A} \|_F$ denotes the Frobenius norm of matrix $\mathbf{A}$.
Finally, we adopt a Matlab-like matrix indexing notation: $\mathbf{A}(:,i)$ denotes the $i$-th column of matrix $\mathbf{A}$; $\mathbf{A}(i,j)$ denotes the element of the $i$-th row and the $j$-th column of  matrix $\mathbf{A}$; $\mathbf{a}(i)$ denotes the $i$-th element of vector $\mathbf{a}$.

\section{System Model and Problem Formulation}
\label{sc:system model}

\subsection{Point-to-Point mmWave MIMO System Model}

\begin{figure*}[!t]
\centering
\vspace{-0.0 cm}
\includegraphics[width= 5.7 in]{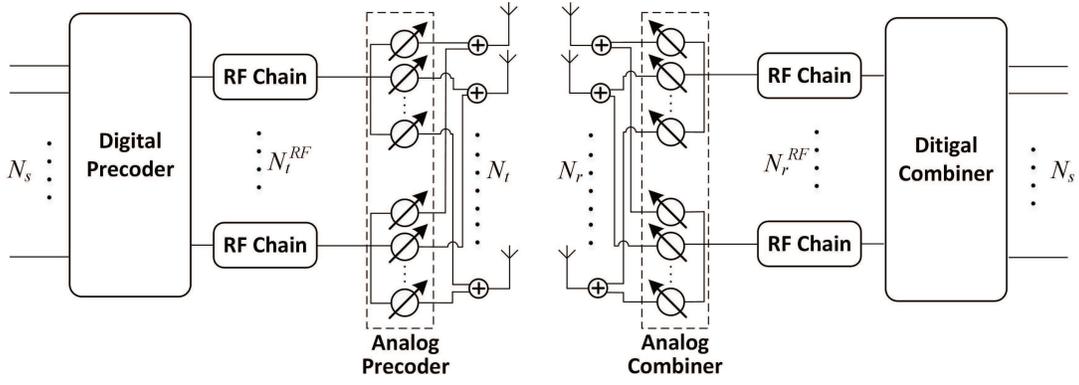}
\caption{The point-to-point mmWave MIMO system using hybrid precoder and combiner.}\label{fig:SU_system_model}\vspace{-0.0 cm}
\end{figure*}

We first consider a point-to-point mmWave MIMO system using a hybrid precoder and combiner with low-resolution PSs, as illustrated in Fig. \ref{fig:SU_system_model}. The transmitter employs $N_t$ antennas and $N^{RF}_t$ RF chains to simultaneously transmit $N_s$ data streams to the receiver which is equipped with $N_r$ antennas and $N^{RF}_r$ RF chains. To ensure the efficiency of the mmWave communication with a limited number of RF chains, the number of data streams and the number of RF chains are constrained as $N_s = N_t^{RF} = N_r^{RF}$.

The transmitted symbols are first processed by a baseband digital precoder $\mathbf{F}_{BB} \in \mathbb{C}^{N_t^{RF}\times N_s}$, then up-converted to the RF domain via $N_t^{RF}$ RF chains before being precoded with an analog precoder $\mathbf{F}_{RF}$ of dimension $ N_t \times N_t^{RF}$. While the baseband digital precoder $\mathbf{F}_{BB}$ enables both amplitude and phase modifications, the elements of the analog precoder $\mathbf{F}_{RF}$, which are implemented by the PSs, have a constant amplitude $\frac{1}{\sqrt{N_t}}$ and quantized phases: $\mathbf{F}_{RF}(i,j) = \frac{1}{\sqrt{N_t}}e^{j\vartheta_{i,j}}$, in which the phase $\vartheta_{i,j}$ is quantized as $\vartheta_{i,j} \in \mathcal{B}\triangleq\{\frac{2\pi b}{2^{B}}\mid b=1,2,\ldots,2^{B}\}$, and $B$ is the number of bits to control the phase. We denote the constraint set of the analog precoder as follows: $\mathbf{F}_{RF}(i,j)\in \mathcal{F} \triangleq \{\frac{1}{\sqrt{N_t}}e^{j\frac{2\pi b}{2^{B}}}\mid b=1,2,\ldots,2^{B}\}$.
Obviously, a larger number of bits $B$ leads to finer resolution for the PSs and potentially better performance, but also results in  higher hardware complexity and power consumption.

The discrete-time transmitted signal can be written in the following form
\begin{equation}
\mathbf{x} = \sqrt{P}\mathbf{F}_{RF} \mathbf{F}_{BB} \mathbf{s}
\end{equation}
where $\mathbf{s}$ is the $N_s \times 1$ symbol vector, $\mathbb{E}\{ \mathbf{s} \mathbf{s}^H \} = \frac{1}{N_s}\mathbf{I}_{N_s}$, $P$ represents transmit power and this power constraint is enforced
by normalizing $\mathbf{F}_{BB}$ such that $\|\mathbf{F}_{RF}\mathbf{F}_{BB}\|^2_F = N_s$.

We consider a narrow-band slow-fading propagation channel, which yields the following received signal
\bea
\mathbf{y} &= & \mathbf{H}  \mathbf{x}  + \mathbf{n} \non \\
&= &\sqrt{P} \mathbf{H} \mathbf{F}_{RF}\mathbf{F}_{BB}\mathbf{s} + \mathbf{n} \label{eq:received signal 1}
\eea
where $\mathbf{y}$ is the $N_r \times 1$ received signal vector, $\mathbf{H}$ is the $N_r \times N_t$ channel matrix, and  $\mathbf{n}\thicksim\mathcal{CN}(\mathbf{0},\sigma^2\mathbf{I}_{N_r})$ is the complex Gaussian noise vector corrupting the received signal.

The receiver employs an analog combiner implemented by the PSs and a digital combiner using $N_r^{RF}$ RF chains to process the received signal. The signal after the spatial processing has the form
\begin{equation}
\mathbf{\widehat{s}} = \sqrt{P}\mathbf{W}^H_{BB}\mathbf{W}^H_{RF} \mathbf{H} \mathbf{F}_{RF}\mathbf{F}_{BB}\mathbf{s} + \mathbf{W}^H_{BB}\mathbf{W}^H_{RF}\mathbf{n} \label{eq:received signal 2}
\end{equation}
where $\mathbf{W}_{RF}$ is the $N_r \times N_r^{RF}$ analog combiner whose elements have the same constraint as $\mathbf{F}_{RF}$, i.e. $\mathbf{W}_{RF}(i,j)= \frac{1}{\sqrt{N_r}}e^{j\varphi_{i,j}}$, $\varphi_{i,j} \in \mathcal{B}$ and thus $\mathbf{W}_{RF}(i,j)\in \mathcal{W} \triangleq \{\frac{1}{\sqrt{N_r}}e^{j\frac{2\pi b}{2^{B}}}\mid b=1,2,\ldots,2^{B}\}$, $\mathbf{W}_{BB}$ is the $N_r^{RF} \times N_s$ digital baseband combiner and the combiner matrices are normalized such that $\|\mathbf{W}_{RF}\mathbf{W}_{BB}\|^2_F = N_s$.

\subsection{Problem Formulation}

We consider the practical and hardware-efficient scenario in which the PSs have very low-resolution (e.g. $B = 1,2$) to reduce the power consumption and complexity. Under this hardware constraint, we aim to jointly design the hybrid precoder and combiner for a mmWave MIMO system.
When Gaussian symbols are transmitted over the mmWave MIMO channel, the achievable spectral efficiency is given by
\begin{eqnarray}
R \hspace{-0.2 cm} & = & \hspace{-0.2 cm} \mathrm{log}_2 \Bigg( \bigg|  \mathbf{I}_{N_s} + \frac{P}{N_s} \mathbf{R}_n^{-1}  \mathbf{W}^H_{BB}\mathbf{W}^H_{RF} \mathbf{H} \mathbf{F}_{RF}\mathbf{F}_{BB} \times   \nonumber \\ & & \hspace{2.2 cm}  \mathbf{F}_{BB}^H \mathbf{F}_{RF}^H \mathbf{H}^H \mathbf{W}_{RF}  \mathbf{W}_{BB}  \bigg| \Bigg), \label{eq:spectral efficiency}
\end{eqnarray}
where $\mathbf{R}_n \triangleq \sigma_n^2 \mathbf{W}^H_{BB}\mathbf{W}^H_{RF}\mathbf{W}_{RF}\mathbf{W}_{BB}$ is the noise covariance matrix after combining. We aim to jointly design the digital beamformers $\mathbf{F}_{BB}$, $\mathbf{W}_{BB}$ as well as the low-resolution analog beamformers $\mathbf{F}_{RF}$, $\mathbf{W}_{RF}$ to maximize the spectral efficiency:
\begin{equation}
\begin{aligned}
\Big\{\mathbf{F}_{RF}^\star, \mathbf{F}_{BB}^\star, \mathbf{W}_{RF}^\star, \mathbf{W}_{BB}^\star\Big\}= \textrm{arg}~\textrm{max} ~R\\
&\hspace{-4.6 cm}\textrm{s. t.}~~~~\mathbf{F}_{RF}(i,j) \in \mathcal{F}, \forall i,j,\\
&\hspace{-3.6 cm}\mathbf{W}_{RF}(i,j) \in \mathcal{W}, \forall i,j,\\
&\hspace{-3.6 cm}\|\mathbf{F}_{RF}\mathbf{F}_{BB}\|^2_F = N_s,\\
&\hspace{-3.6 cm}\|\mathbf{W}_{RF}\mathbf{W}_{BB}\|^2_F = N_s.
\end{aligned}\label{eq:optimization problem}
\end{equation}
Obviously, the optimization problem (\ref{eq:optimization problem}) is a non-convex NP-hard problem. In the next section, we attempt to decompose the original problem into a series of sub-problems and seek a sub-optimal solution with low-complexity and satisfactory performance.

\section{Low-Resolution Hybrid Precoder and Combiner Design}
\label{sec:Proposed Design Full Connect}

To simplify the joint hybrid precoder and combiner design, the objective problem is decomposed into two
separate optimizations. We first focus on the joint design of the analog precoder $\mathbf{F}_{RF}$ and combiner $\mathbf{W}_{RF}$. Then, having the effective baseband channel associated with the obtained optimal analog precoder and combiner, the digital precoder $\mathbf{F}_{BB}$ and combiner $\mathbf{W}_{BB}$ are computed to further maximize the spectral efficiency.

\subsection{Low-Resolution Analog Precoder and Combiner Design}
\label{sec:low resolution analog beamformer}

We observe that under the assumption of high signal-to-noise-ratio (SNR), the achievable spectral efficiency in (\ref{eq:spectral efficiency}) can be approximated as
\begin{eqnarray}
R \hspace{-0.2 cm} & \approx & \hspace{-0.2 cm} \mathrm{log}_2 \Bigg( \bigg|  \frac{P}{N_s} \mathbf{R}_n^{-1}  \mathbf{W}^H_{BB}\mathbf{W}^H_{RF} \mathbf{H} \mathbf{F}_{RF}\mathbf{F}_{BB} \times   \nonumber \\ & & \hspace{1.6 cm}  \mathbf{F}_{BB}^H \mathbf{F}_{RF}^H \mathbf{H}^H \mathbf{W}_{RF}  \mathbf{W}_{BB}  \bigg| \Bigg).\label{eq:high SNR}
\end{eqnarray}
While the per-antenna SNR in mmWave systems is typically low, the post-combining SNR should be high enough to justify this approximation. In addition, it has been verified in \cite{Sohrabi 16} that for large-scale MIMO systems, the optimal analog beamformers are approximately orthogonal, i.e. $\mathbf{F}_{RF}^H\mathbf{F}_{RF} \propto \mathbf{I}_{N^{RF}_t}$. This enables us to assume $\mathbf{F}_{BB}\mathbf{F}_{BB}^H\approx\zeta^2\mathbf{I}_{N_s}$ when $N^{RF}_t = N_s$, where $\zeta^2$ is a normalization factor. Similarly, we have $\mathbf{W}_{BB}\mathbf{W}_{BB}^H\approx\xi^2\mathbf{I}_{N_s}$ and $\mathbf{W}_{BB}^H\mathbf{W}_{RF}^H\mathbf{W}_{RF}\mathbf{W}_{BB}\approx\mathbf{I}_{N_s}$. Let $\gamma^2\triangleq\zeta^2\xi^2$, then (\ref{eq:high SNR}) can be further simplified as
\begin{align}
&\hspace{-0.3 cm}R\approx \mathrm{log}_2 \Bigg( \bigg|  \frac{P\gamma^2}{N_s\sigma^2}\mathbf{W}^H_{RF} \mathbf{H} \mathbf{F}_{RF} \mathbf{F}_{RF}^H \mathbf{H}^H \mathbf{W}_{RF} \bigg| \Bigg)\\
&\hspace{-0.1 cm}\overset{(a)}{=}N_s\mathrm{log}_2 \left(\frac{P\gamma^2}{N_s\sigma^2}\right)+2\times \mathrm{log}_2 \Bigg( \bigg|\mathbf{W}^H_{RF} \mathbf{H} \mathbf{F}_{RF}\bigg| \Bigg)
\end{align}
where $(a)$ follows since $|\mathbf{X}\mathbf{Y}| = |\mathbf{X}||\mathbf{Y}|$ when $\mathbf{X}$ and $\mathbf{Y}$ are both square matrices. Therefore, the analog precoder and combiner design with low-resolution PSs can be approximately reformulated as:
\begin{equation}
\begin{aligned}
\left\{\mathbf{F}_{RF}^\star, \mathbf{W}_{RF}^\star \right\}= & \; \textrm{arg} \; {\textrm{max}} \; \mathrm{log}_2 \bigg( \Big|\mathbf{W}^H_{RF} \mathbf{H} \mathbf{F}_{RF}\Big| \bigg)  \\
\mathrm{s. \, t. } \; & \mathbf{F}_{RF}(i,j) \in \mathcal{F}, \forall i,j, \\ & \mathbf{W}_{RF}(i,j) \in \mathcal{W}, \forall i,j.
\end{aligned}\label{eq:objective function}
\end{equation}
Unfortunately, the optimization problem (\ref{eq:objective function}) is still NP-hard and has exponential complexity $\mathcal{O}(|\mathcal{F}|^{N_t N_t^{RF}} |\mathcal{W}|^{ N_r N_r^{RF}})$. Therefore, we propose to further decompose this difficult optimization problem into a series of sub-problems, in which each transmit/receive RF chain pair is considered one by one and the analog precoder and combiner for each pair are successively designed.

In particular, we define the singular value decomposition (SVD) of $\mathbf{H}$ as
\begin{equation}
\mathbf{H}=\mathbf{U}\mathbf{\Sigma}\mathbf{V}^H
\end{equation}
where $\mathbf{U}$ is an $N_r \times N_r$ unitary matrix, $\mathbf{V}$ is an $N_t \times N_t$ unitary matrix, and $\mathbf{\Sigma}$ is a rectangular diagonal matrix of singular values. Due to the sparse nature of the mmWave channel, the matrix $\mathbf{H}$ is typically low rank. In particular, the effective rank of the channel serves as an upper bound for the number of data streams $N_s$ that the channel can support. Thus, we assume that the channel $\mathbf{H}$ can be well approximated by retaining only the $N_s$ strongest components 
$\mathbf{H} \approx \mathbf{\widehat{U}}\mathbf{\widehat{\Sigma}}\mathbf{\widehat{V}}^H$, where
$\mathbf{\widehat{U}}\triangleq\mathbf{U}(:,1:N_s)$, $\mathbf{\widehat{\Sigma}}\triangleq\mathbf{\Sigma}(1:N_s,1:N_s)$, and $\mathbf{\widehat{V}}\triangleq\mathbf{V}(:,1:N_s)$.
Then, the objective in (\ref{eq:objective function}) can be converted to
\begin{eqnarray}
&&\hspace{-1.0 cm}\mathrm{log}_2 \bigg( \Big|\mathbf{W}^H_{RF} \mathbf{H} \mathbf{F}_{RF}\Big| \bigg)
\approx\mathrm{log}_2 \bigg( \Big|\mathbf{W}^H_{RF} \mathbf{\widehat{U}}\mathbf{\widehat{\Sigma}}\mathbf{\widehat{V}}^H \mathbf{F}_{RF}\Big| \bigg).\label{eq:decomposition}
\end{eqnarray}

Next, we write the analog precoding and combining matrices as $\mathbf{F}_{RF}\triangleq[\mathbf{f}_{RF,{1}}   \ldots \mathbf{f}_{RF,{N_s}}]$ and $\mathbf{W}_{RF}\triangleq[\mathbf{w}_{RF,{1}} \ldots \mathbf{w}_{RF,{N_s}}]$, respectively, where $\mathbf{f}_{RF,l}$ and $\mathbf{w}_{RF,l}$, $l=1,\ldots, N_s$, are the analog precoder and combiner pair for the $l$-th data stream.
Furthermore, we denote $\mathbf{F}_{RF, {\backslash l}}$ as the precoding matrix excluding the $l$-th precoder vector $\mathbf{f}_{RF,l}$ and $\mathbf{W}_{RF, {\backslash l}}$ as the combining matrix excluding the $l$-th combiner vector $\mathbf{w}_{RF,l}$.
Then, the formulation (\ref{eq:decomposition}) can be further transformed to (\ref{eq:decomposition1})-(\ref{eq:transformation}), which are presented at the top of following page,
\begin{figure*}
\begin{small}
\begin{eqnarray}
&&\hspace{-1.0 cm}\mathrm{log}_2 \bigg( \Big|\mathbf{W}^H_{RF} \mathbf{\widehat{U}}\mathbf{\widehat{\Sigma}}\mathbf{\widehat{V}}^H \mathbf{F}_{RF}\Big| \bigg)=\mathrm{log}_2 \bigg( \Big|\mathbf{\widehat{\Sigma}}\mathbf{\widehat{V}}^H \mathbf{F}_{RF}\mathbf{W}^H_{RF} \mathbf{\widehat{U}}\Big| \bigg) \label{eq:decomposition1} \\
&&\hspace{-0.8 cm} =\mathrm{log}_2 \Bigg( \bigg|\mathbf{\widehat{\Sigma}}\mathbf{\widehat{V}} ^H \left[\mathbf{F}_{RF,{\backslash l}}~\mathbf{f}_{RF,{l}}\right]\left[\mathbf{W}_{RF,{\backslash l}}~ \mathbf{w}_{RF,l}\right]^H \mathbf{\widehat{U}}\bigg| \Bigg)
=\mathrm{log}_2 \Bigg( \bigg|\mathbf{\widehat{\Sigma}}\mathbf{\widehat{V}}^H \mathbf{F}_{RF,{\backslash l}}\mathbf{W}_{RF,{\backslash l}}^H\mathbf{\widehat{U}}+ \mathbf{\widehat{\Sigma}}\mathbf{\widehat{V}}^H\mathbf{f}_{RF,l}\mathbf{w}_{RF,l}^H \mathbf{\widehat{U}} \bigg| \Bigg)\\
&&\hspace{-0.8 cm}\approx\mathrm{log}_2 \Bigg( \bigg|\left(\mathbf{\widehat{\Sigma}}\mathbf{\widehat{V}}^H \mathbf{F}_{RF,{\backslash l}}\mathbf{W}_{RF,{\backslash l}}^H\mathbf{\widehat{U}}\right)\Big[ \mathbf{I}_{N_s}+\left(\alpha\mathbf{I}_{N_s}+\mathbf{\widehat{\Sigma}}\mathbf{\widehat{V}}^H \mathbf{F}_{RF,{\backslash l}}\mathbf{W}_{RF,{\backslash l}}^H\mathbf{\widehat{U}}\right)^{-1} \mathbf{\widehat{\Sigma}}\mathbf{\widehat{V}}^H\mathbf{f}_{RF,l}\mathbf{w}_{RF,l}^H \mathbf{\widehat{U}}\Big] \bigg| \Bigg)\\
&&\hspace{-0.8 cm}=\mathrm{log}_2 \Bigg( \bigg| \mathbf{\widehat{\Sigma}}\mathbf{\widehat{V}}^H \mathbf{F}_{RF,{\backslash l}}\mathbf{W}_{RF,{\backslash l}}^H\mathbf{\widehat{U}} \bigg| \Bigg)+
\mathrm{log}_2 \Bigg( \bigg|\Big[ \mathbf{I}_{N_s}+\left(\alpha\mathbf{I}_{N_s}+\mathbf{\widehat{\Sigma}}\mathbf{\widehat{V}}^H \mathbf{F}_{RF,{\backslash l}}\mathbf{W}_{RF,{\backslash l}}^H\mathbf{\widehat{U}}\right)^{-1} \mathbf{\widehat{\Sigma}}\mathbf{\widehat{V}}^H\mathbf{f}_{RF,l}\mathbf{w}_{RF,l}^H \mathbf{\widehat{U}}\Big] \bigg| \Bigg) \\
&&\hspace{-0.8 cm}=\mathrm{log}_2 \Bigg( \bigg|  \mathbf{W}_{RF,{\backslash l}}^H \mathbf{\widehat{U}} \mathbf{\widehat{\Sigma}}\mathbf{\widehat{V}}^H \mathbf{F}_{RF,{\backslash l}}  \bigg| \Bigg)+
\mathrm{log}_2 \Bigg( \bigg|\Big[ 1+\mathbf{w}_{RF,l}^H\mathbf{\widehat{U}}\left(\alpha\mathbf{I}_{N_s}+\mathbf{\widehat{\Sigma}}\mathbf{\widehat{V}}^H \mathbf{F}_{RF,{\backslash l}}\mathbf{W}_{RF,{\backslash l}}^H\mathbf{\widehat{U}}\right)^{-1} \mathbf{\widehat{\Sigma}}\mathbf{\widehat{V}}^H\mathbf{f}_{RF,l}\Big] \bigg| \Bigg)  \label{eq:transformation}
\end{eqnarray}
\end{small}
\hrule \vspace{-0.0 cm}
\end{figure*}
where $\alpha$ is a very small scalar to assure invertibility. Thus, the objective in (\ref{eq:objective function}) can be reformulated as:
\begin{eqnarray}
&&\hspace{-1 cm}\mathrm{log}_2 \bigg( \Big|\mathbf{W}^H_{RF} \mathbf{H} \mathbf{F}_{RF}\Big| \bigg) \approx \mathrm{log}_2 \Bigg( \bigg|   \mathbf{W}_{RF,{\backslash l}}^H \mathbf{H}  \mathbf{F}_{RF,{\backslash l}}  \bigg| \Bigg) \nonumber \\ && \hspace{2.0 cm} + \mathrm{log}_2 \Bigg( \bigg| \mathbf{w}_{RF,l}^H \mathbf{Q}_l\mathbf{f}_{RF,l} \bigg| \Bigg) \label{eq:final}
\end{eqnarray}
\nid where we define the interference-included channel matrix $ \mathbf{Q}_l$ as
\be \mathbf{Q}_l \triangleq \mathbf{\widehat{U}}(\alpha\mathbf{I}_{N_s}+\mathbf{\widehat{\Sigma}}\mathbf{\widehat{V}}^H \mathbf{F}_{RF,{\backslash l}} \mathbf{W}_{RF,{\backslash l}}^H\mathbf{\widehat{U}})^{-1} \mathbf{\widehat{\Sigma}}\mathbf{\widehat{V}}^H . \label{eq:n1}\ee
According to (\ref{eq:final}), if $\mathbf{F}_{RF,{\backslash l}}$ and $\mathbf{W}_{RF,{\backslash l}}$ are known, the problem (\ref{eq:objective function}) can be reformulated as finding a corresponding precoder $\mathbf{f}_{RF,l}$ and combiner $\mathbf{w}_{RF,l}$ pair to conditionally maximize the achievable spectral efficiency:
\begin{equation}
\begin{aligned}
 \left\{\mathbf{f}_{{RF},l}^{\star},\mathbf{w}_{{RF},l}^{\star}  \right\}= & \; \textrm{arg}
\; \textrm{max} \; \left| \mathbf{w}_{RF,l}^H \mathbf{Q}_l\mathbf{f}_{RF,l} \right|  \\
\mathrm{s. \; t. \;\; }&  \mathbf{f}_{{RF},l}(i)  \in \mathcal{F}, ~ i=1,\ldots, N_t,\\
& \mathbf{w}_{{RF},l}(j)  \in \mathcal{W}, ~ j=1,\ldots, N_r.
\label{eq:beam sel}
\end{aligned}
\end{equation}
\nid This motivates us to propose an iterative algorithm, which starts with
appropriate initial RF precoding and combining matrices then successively designs $\mathbf{f}_{{RF},l}$ and $\mathbf{w}_{{RF},l}$ according to (\ref{eq:beam sel}) with an updated $\mathbf{Q}_l$ as in (\ref{eq:n1}) until the algorithm converges.

The complexity of obtaining an optimal solution to (\ref{eq:beam sel}) for each iteration is now reduced to $\mathcal{O}(|\mathcal{F}|^{N_t} |\mathcal{W}|^{ N_r })$, which is still too high. To practically solve the problem (\ref{eq:beam sel}), in what follows we present an iterative phase matching algorithm, which searches the conditionally optimal phase of each element of the analog precoder $\mathbf{f}_{{RF},l}$ and combiner $\mathbf{w}_{{RF},l}$. Specifically, we first design the analog precoder $\mathbf{f}_{{RF},l}$ assuming the analog combiner $\mathbf{w}_{RF,l}$ is fixed. Let $\vartheta_{l,i}$ be the phase of the $i$-th element of the analog precoder $\mathbf{f}_{{RF},l}$ and let $\varphi_{l,j}$ be the phase of the $j$-th element of the analog combiner $\mathbf{w}_{RF,l}$. If we temporarily remove the discrete  phase constraint, the optimal continuous phase $\tilde{\vartheta}_{l,i}$ of the $i$-th element of the analog precoder $\mathbf{f}_{RF,l}$ is given by the following proposition, whose proof is provided in Appendix A.

\emph{Proposition 1:} Given the phases $\varphi_{l,j}$ of the analog combiner $\mathbf{w}_{RF,l}$ and the phases $\vartheta_{l,u}$, $u \neq i$, of the analog precoder $\mathbf{f}_{RF,l}$, the optimal continuous phase $\tilde{\vartheta}_{l,i}$ of the $i$-th element of analog precoder $\mathbf{f}_{RF,l}$ is
\bea
\hspace{0.4 cm}\tilde{\vartheta}_{l,i}=\mathrm{angle} \left\{\sum_{j=1}^{N_r}e^{j\varphi_{l,j}} \sum_{u\neq i}^{N_t}e^{j\vartheta_{l,u}}\mathbf{Q}_l(j,u)\right\} \non \\
&&\hspace{-4.5 cm}-\mathrm{angle} \left\{\sum_{j=1}^{N_r}e^{j\varphi_{l,j}} \mathbf{Q}_l(j,i)\right\}.\label{eq:opt phase f}
\eea   $\hfill$ $\blacksquare$

\nid Then, after finding the optimal continuous phase $\tilde{\vartheta}_{l,i}$ by (\ref{eq:opt phase f}), we reconsider the discrete phase constraint and find the optimal low-resolution phase $\vartheta_{l,i}$ by quantization:
\be
\vartheta_{l,i}=\textrm{arg}\underset{\substack{ \hat{\vartheta}_{l,i} \in \mathcal{B}}}
 {\textrm{min}} \big| \tilde{\vartheta}_{l,i}-\hat{\vartheta}_{l,i} \big|. \label{eq:opt qt phase f}
\ee

\nid Similarly, if the analog precoder $\mathbf{f}_{RF,l}$ is determined, the optimal continuous phase $\tilde{\varphi}_{l,j}$ of the $j$-th element of $\mathbf{w}_{RF,l}$ is
\bea
\hspace{0.4 cm}\tilde{\varphi}_{l,j} = \mathrm{angle} \left\{\sum_{i=1}^{N_t}e^{j\vartheta_{l,i}} \sum_{u\neq j}^{N_t}e^{j\varphi_{l,u}}\mathbf{Q}_l(u,i)\right\} \non \\ && \hspace{-4.5 cm} -\mathrm{angle} \left\{\sum_{i=1}^{N_t}e^{j\varphi_{l,i}} \mathbf{Q}_l(j,i)\right\}, \label{eq:opt phase w}
\eea
and the optimal low-resolution phase $\varphi_{l,j}$ is obtained by
\bea
\varphi_{l,j}=\textrm{arg}\underset{\substack{  \hat{\varphi}_{l,j}  \in \mathcal{B}}}
 {\textrm{min}} \big| \tilde{\varphi}_{l,j} -\hat{\varphi}_{l,j} \big|.\label{eq:opt qt phase w}
\eea

Motivated by (\ref{eq:opt phase f})-(\ref{eq:opt qt phase w}), the iterative procedure to design the precoder $\mathbf{f}_{{RF},l}$ and combiner $\mathbf{w}_{{RF},l}$ as in (\ref{eq:beam sel}) is straightforward. With appropriate initial ${\vartheta}_{l,i}$,  $\varphi_{l,j}$, we design the precoder $\mathbf{f}_{{RF},l}$ by finding the conditionally optimal phases $\vartheta_{l,i}$ as in (\ref{eq:opt phase f}) and (\ref{eq:opt qt phase f}). Then, with the obtained $\vartheta_{l,i}$, $i=1,\ldots, N_t$, we design the combiner $\mathbf{w}_{{RF},l}$ by finding the conditionally optimal phases $\varphi_{l,j}$  as in (\ref{eq:opt phase w}) and (\ref{eq:opt qt phase w}).
We alternate the designs of $\mathbf{f}_{{RF},l}$ and $\mathbf{w}_{{RF},l}$ iteratively until the obtained phase of each element of $\mathbf{f}_{{RF},l}$ and $\mathbf{w}_{{RF},l}$ does not change and the convergence is achieved.
Note that since in each precoder and combiner design step, the objective
function of (\ref{eq:beam sel}) is monotonically non-decreasing, and thus our proposed algorithm is guaranteed to converge to at least a locally optimal solution.

We summarize the proposed joint low-resolution analog precoder and combiner design in Algorithm \ref{alg:phase matching}.

\begin{algorithm}[htb!]
\caption {Iterative Phase Matching Algorithm for Low-Resolution Analog Precoder and Combiner Design}
\label{alg:phase matching}
\begin{algorithmic}[1]
\REQUIRE  $\mathcal{F}$, $\mathcal{W}$, $\mathbf{H}$.
\ENSURE  $\mathbf{F}_{RF}^\star$ and $\mathbf{W}_{RF}^\star$.
\STATE \hspace{-0.1 cm}Initialize  $\mathbf{F}_{RF}^\star = \mathbf{0}$, $\mathbf{W}_{RF}^\star = \mathbf{0}$.
\FOR {$l=1:N_s$}
\STATE Obtain $\mathbf{F}_{RF,{\backslash l}}$ from $\mathbf{F}_{RF}^\star$ and $\mathbf{W}_{RF,{\backslash l}}$ from $\mathbf{W}_{RF}^\star$.
\STATE Update $\mathbf{Q}_l \hspace*{-0.1 cm} = \hspace*{-0.1 cm} \mathbf{\widehat{U}}(\alpha\mathbf{I}_{N_s} \hspace*{-0.1 cm} + \hspace*{-0.05 cm} \mathbf{\widehat{\Sigma}}\mathbf{\widehat{V}}^H\mathbf{F}_{RF,{\backslash l}}\mathbf{W}_{RF,{\backslash l}}^H\mathbf{\widehat{U}})^{-1}\mathbf{\widehat{\Sigma}}\mathbf{\widehat{V}}^H $.
\WHILE {no convergence of ${\vartheta}_{l,i}$ and $\varphi_{l,j}$}
\FOR {$i=1:N_t$}
\STATE Obtain quantized phase $\vartheta_{l,i}$ by (\ref{eq:opt phase f}) and (\ref{eq:opt qt phase f}).
\ENDFOR
\FOR {$j=1:N_r$}
\STATE Obtain quantized phase $\varphi_{l,j}$ by (\ref{eq:opt phase w}) and (\ref{eq:opt qt phase w}).
\ENDFOR
\ENDWHILE
\STATE Construct $\mathbf{f}_{RF,l}^\star$ by $\vartheta_{l,i}$ and $\mathbf{w}_{RF,l}^\star$ by $\varphi_{l,j}$.
\ENDFOR
\STATE Construct $\mathbf{F}_{RF}^\star$ by $\mathbf{f}_{RF,l}^\star$  and $\mathbf{W}_{RF}^\star$ by $\mathbf{w}_{RF,l}^\star$.
\STATE Goto Step 2 until convergence  of $\mathbf{F}_{RF}^\star$ and $\mathbf{W}_{RF}^\star$.
\end{algorithmic}
\end{algorithm}

\subsection{Digital Precoder and Combiner Design}

After all analog precoder-combiner pairs have been determined, we can obtain the effective baseband channel $\mathbf{\widetilde{H}}$ as
\begin{equation}
\mathbf{\widetilde{H}} \triangleq \left(\mathbf{W}_{RF}^\star\right)^H \mathbf{H} \mathbf{F}_{RF}^\star,  \label{eq:effective channel}
\end{equation}
where $\mathbf{F}_{RF}^\star \triangleq [\mathbf{f}_{RF,1}^\star, \ldots, \mathbf{f}_{RF,{N_s}}^\star] $ and $\mathbf{W}_{RF}^\star\triangleq [\mathbf{w}_{RF,1}^\star, \ldots, \mathbf{w}_{RF,{N_s}}^\star]$. For the baseband precoder and combiner design, we define the SVD of the effective baseband channel $\mathbf{\widetilde{H}}$ as
\begin{equation}
\mathbf{\widetilde{H}} = \mathbf{\widetilde{U}} \mathbf{\widetilde{\Sigma}} \mathbf{\widetilde{V}}^H
\end{equation}
where $ \mathbf{\widetilde{U}}$ and $\mathbf{\widetilde{V}}$ are $N_s \times N_s$ unitary matrices, $\mathbf{\widetilde{\Sigma}}$ is an $N_s \times N_s$ diagonal matrix of singular values. Then, to further enhance the spectral efficiency, an SVD-based baseband digital precoder and combiner are employed:
\begin{eqnarray}
&\mathbf{F}^\star_{BB}  = \mathbf{\widetilde{V}}, \\
&\mathbf{W}^\star_{BB}  =  \mathbf{\widetilde{U}}.
\end{eqnarray}
Finally, the baseband precoder and combiner are normalized
\begin{eqnarray}
&\mathbf{F}^\star_{{BB}}=\frac{\sqrt{N_s}\mathbf{F}^\star_{{BB}}} {\|\mathbf{F}^\star_{RF}\mathbf{F}^\star_{{BB}}\|_F},\\
&\mathbf{W}^\star_{{BB}}=\frac{\sqrt{N_s}\mathbf{W}^\star_{{BB}}} {\|\mathbf{W}^\star_{RF}\mathbf{W}^\star_{{BB}}\|_F}.
\end{eqnarray}

\section{One-Bit Resolution Analog Precoder and Combiner Design}
\label{sec:low complexity scheme}

In the previous section, we proposed a novel hybrid beamformer design for maximizing the spectral efficiency of a mmWave MIMO system, in which the analog precoder and combiner are implemented with low-resolution PSs.
In order to achieve maximum hardware efficiency, in this section we focus on the design of analog precoders and combiners using ``one-bit'' resolution (binary) PSs, which can maximally reduce the power consumption and simplify the hardware complexity.
Although the iterative phase matching algorithm proposed in the previous section can also be applied, a simpler approach is possible in the one-bit case. Therefore, in this section, we present an efficient one-bit resolution analog beamformer design, which can achieve good performance with much lower complexity.

We follow the procedure of the hybrid beamforming design proposed in the previous section, but only modify the optimization problem (\ref{eq:beam sel}), which attempts to determine the $l$-th analog precoder and combiner pair.
Particularly, we reformulate this analog beamformer design problem (\ref{eq:beam sel}) with the constraint of one-bit resolution PSs as
\begin{equation}
\left\{\mathbf{f}_{{RF},l}^{\star},\mathbf{w}_{{RF},l}^{\star}  \right\}= \textrm{arg}\underset{\substack{\mathbf{f}_{{RF},l}  \in \frac{1}{\sqrt{N_t}}\{\pm 1\}^{N_t}\\ \mathbf{w}_{{RF},l}  \in \frac{1}{\sqrt{N_r}}\{\pm 1\}^{N_r}}}
 {\textrm{max}} \left| \mathbf{w}_{RF,l}^H \mathbf{Q}_l\mathbf{f}_{RF,l} \right|.
\label{eq:beam sel_1-bit}
\end{equation}
The optimization problem (\ref{eq:beam sel_1-bit}) can be solved through exhaustive search with exponential complexity $\mathcal{O}(2^{N_t N_r})$, which would not be possible with large antenna arrays. Therefore, in the following we attempt to develop an efficient one-bit resolution beamformer design with polynomial complexity in the number of antennas.

We first define the SVD of $\mathbf{Q}_l$ as
\begin{equation}
\mathbf{Q}_l=\sum_{i=1}^{N_s}\lambda_{l,i}\mathbf{p}_{l,i} \mathbf{g}_{l,i}^H, \label{eq:Q_l SVD1}
\end{equation}
where $\mathbf{p}_{l,i}$ and $\mathbf{g}_{l,i}$ are the $i$-th left and right singular vectors of $\mathbf{Q}_l$, respectively, and $\lambda_{l,i}$ is the $i$-th largest singular value, $\lambda_{l,1}\geq\lambda_{l,2} \geq\ldots\geq\lambda_{l,{N_s}}$. Then, the objective in (\ref{eq:beam sel_1-bit}) can be rewritten as
\be | \mathbf{w}_{RF,l}^H \mathbf{Q}_l\mathbf{f}_{RF,l} |= \left| \sum_{i=1}^{N_s }\lambda_{l,i}\mathbf{w}_{RF,l}^H \mathbf{p}_{l,i} \mathbf{g}_{l,i}^H\mathbf{f}_{RF,l} \right|. \label{eq:Q_l SVD}\ee
If we utilize a rank-1 approximation by keeping only the strongest term, i.e.  $\mathbf{Q}_l\approx\lambda_{l,1}\mathbf{p}_{l,1}\mathbf{g}_{l,1}^H$, the optimization function in (\ref{eq:beam sel_1-bit}) can be approximated by
\begin{equation}
\begin{aligned}
\hspace{-0.1 cm}\left\{\mathbf{f}_{{RF},{l}}^{\star},\mathbf{w}_{{RF},{l}}^\star  \right\}= \textrm{arg}\hspace{-0.3 cm}\underset{\substack{\mathbf{f}_{{RF},l}  \in \frac{1}{\sqrt{N_t}}\{\pm 1\}^{N_t}\\ \mathbf{w}_{{RF},l}  \in \frac{1}{\sqrt{N_r}}\{\pm 1\}^{N_r}}}{\textrm{max}}\hspace{-0.2 cm} \left| \mathbf{w}_{RF,l}^H \mathbf{p}_{l,1} \mathbf{g}_{l,1}^H\mathbf{f}_{RF,l} \right|.
\end{aligned}\label{eq:joint optimize}
\end{equation}
Now, the joint optimization problem (\ref{eq:joint optimize}) can be decoupled into individually designing the analog precoder $\mathbf{f}_{{RF},{l}}$ and combiner  $\mathbf{w}_{{RF},{l}}$:
\begin{eqnarray}
&\mathbf{f}_{{RF},{l}}^{\star} = \textrm{arg}  \underset{\mathbf{f}_{{RF},l}  \in \frac{1}{\sqrt{N_t}}\{\pm1\}^{N_t}}{\textrm{max}}  \left| \mathbf{f}_{RF,l}^H \mathbf{g}_{l,1} \right|,\label{eq:analog precoder design1}\\
&\mathbf{w}_{{RF},{l}}^{\star} = \textrm{arg}  \underset{\mathbf{w}_{{RF},l}  \in \frac{1}{\sqrt{N_r}}\{\pm1\}^{N_r}}{\textrm{max}}  \left| \mathbf{w}_{RF,l}^H \mathbf{p}_{l,1} \right|. \label{eq:analog combiner design1}
\end{eqnarray}
These two optimization problems (\ref{eq:analog precoder design1}) and  (\ref{eq:analog combiner design1}) require only the singular vectors  $\mathbf{p}_{l,1} $ and $\mathbf{g}_{l,1} $ associated with the largest singular value, which can be quickly obtained by the power iteration algorithm \cite{Matrix Computations} instead of the complete SVD calculation. However, solving (\ref{eq:analog precoder design1}) and  (\ref{eq:analog combiner design1}) by exhaustive search still has exponential complexity in the number of antennas.
In order to further reduce the complexity without a significant loss of performance, we propose to construct a smaller dimension candidate beamformer set, from which the optimal beamformer can be found with linear complexity. In the following, we present this algorithm for the precoder design (\ref{eq:analog precoder design1}) as an example, while the combiner design (\ref{eq:analog combiner design1}) follows the same procedure.

We introduce an auxiliary variable $\phi \in [-\pi,\pi)$ and we reformulate the optimization problem (\ref{eq:analog precoder design1}) as:
\bea
&\hspace{-0.5 cm}\left\{\phi^\star, \mathbf{f}_{RF,l}^\star\right\}=\textrm{arg} \hspace{-0.2 cm} \underset{\substack{\phi \in [-\pi,\pi)\\ \mathbf{f}_{{RF},l}  \in \frac{1}{\sqrt{N_t}}\{\pm1\}^{N_t}}}{\textrm{max}}   \hspace{-0.2 cm} \mathfrak{Re}\left\{\mathbf{f}_{RF,l}^H \mathbf{g}_{l,1} e^{-j\phi}\right\} \\
&\hspace{-0.4 cm}=\textrm{arg}\hspace{-0.2 cm}\underset{\substack{\phi \in [-\pi,\pi)\\ \mathbf{f}_{{RF},l}  \in \frac{1}{\sqrt{N_t}}\{\pm1\}^{N_t}}}{\textrm{max}}   \hspace{-0.2 cm} \sum\limits_{i=1}^{N_t}\mathbf{f}_{RF,l}(i) |\mathbf{g}_{l,1}(i)| \cos(\phi-\psi_i)
\label{eq:analog precoder design2}
\eea
where $\psi_i$ denotes the phase of $\mathbf{g}_{l,1}(i)$. Obviously, given any $\phi \in [-\pi,\pi)$, the corresponding binary precoder that maximizes (\ref{eq:analog precoder design2}) is \be \mathbf{f}_{RF,l}(i)=\frac{1}{\sqrt{N_t}}\mathrm{sign}\left(\cos\left(\phi -\psi_i\right)\right), i=1, \ldots , N_t. \label{eq:analog precoder design5} \ee
With the conditionally optimal $\mathbf{f}_{RF,l}$ for any given $\phi$ shown in (\ref{eq:analog precoder design5}), we will now show that we can always construct a set of $N_t$ candidate binary precoders $\mathcal{F}_l \triangleq \{\mathbf{f}_{l,1}, \ldots, \mathbf{f}_{l,{N_t}}\} $ and guarantee $ \mathbf{f}_{{RF},{l}}^{\star} \in \mathcal{F}_l$. Then, the maximization in (\ref{eq:analog precoder design1}) can be carried out over a set of only $N_t$ candidates without loss of performance.

We first define the angles $\widehat{\psi_i}$, $i=1, \ldots, N_t$, as
\begin{equation}
\widehat{\psi_i}\triangleq\left\{
\begin{aligned}
 & \psi_i-\pi, ~~ \mathrm{if} ~ \psi_i \in \left[\frac{\pi}{2},\frac{3\pi}{2}\right), \\
  &\psi_i,~~~~~~~\mathrm{if} ~ \psi_i \in \left[-\frac{\pi}{2},\frac{\pi}{2}\right),
\end{aligned}
\right. \label{eq:psi_i}
\end{equation}
so that $\widehat{\psi_i}\in[-\frac{\pi}{2},\frac{\pi}{2})$.
Then, we map the angles $\widehat{\psi}_i$ to $\widetilde{\psi}_i$, $i=1, \ldots , N_t$, which are rearranged in ascending order, i.e. $\widetilde{\psi}_1 \leq \widetilde{\psi}_2 \leq \ldots \leq \widetilde{\psi}_{N_t} $.
Because of the periodicity of the cosine function, the maximization problem (\ref{eq:analog precoder design2}) with respect to $\phi$
can be carried out over any interval of length $\pi$.
If we construct $N_t$ non-overlapping sub-intervals $[\widetilde{\psi}_1-\frac{\pi}{2}, \widetilde{\psi}_2-\frac{\pi}{2}),[\widetilde{\psi}_2-\frac{\pi}{2}, \widetilde{\psi}_3-\frac{\pi}{2}),\ldots,[\widetilde{\psi}_{N_t}-\frac{\pi}{2}, \widetilde{\psi}_1+\frac{\pi}{2})$, then  the optimal $\phi^\star$ must be located in one of $N_t$ sub-intervals since the full interval is $[\widetilde{\psi}_1-\frac{\pi}{2},\widetilde{\psi}_1+\frac{\pi}{2})$ of length $\pi$. Therefore, the optimization problem (\ref{eq:analog precoder design2}) can be solved by examining each sub-interval separately.

Assuming the optimal $\phi^\star$ is in the $k$-th sub-interval, the corresponding optimal binary precoder can be obtained by (\ref{eq:analog precoder design5}) as $\mathbf{\widetilde{f}}_{l,k}(i) = \frac{1}{\sqrt{N_t}}\mathrm{sign}\left(\cos\left(\phi^\star - \widetilde{\psi}_i\right)\right) $, $i=1,\ldots, N_t$, and has the form
\begin{equation}
\mathbf{\widetilde{f}}_{l,k}= \frac{1}{\sqrt{N_t}}[\underbrace{1 \ldots 1}_k\underbrace{-1 \ldots -1}_{N_t-k}]^T. \label{eq:f_tilde}
\end{equation}
After that, given the inverse sorting that maps $\widetilde{\psi}_i$ to $\widehat{\psi}_i$, we rearrange the corresponding elements of
$\mathbf{\widetilde{f}}_{l,k}$ and obtain $\mathbf{\widehat{f}}_{l,k}$. Then, based on the relationship between $\psi_i$ and $\widehat{\psi}_i$ defined in (\ref{eq:psi_i}), we can achieve the conditionally optimal precoder $\mathbf{f}_{l,k}$ by
\begin{equation}
\begin{aligned}
\hspace{-0.1 cm} \mathbf{f}_{l,k}(i)\triangleq\left\{
\begin{aligned}
&-\mathbf{\widehat{f}}_{l,k}(i), ~\mathrm{if} ~\psi_i \in \left[\frac{\pi}{2},\frac{3\pi}{2}\right),  i=1,\ldots, N_t , \\
&~~\mathbf{\widehat{f}}_{l,k}(i),~~\mathrm{if} ~\psi_i \in \left[-\frac{\pi}{2},\frac{\pi}{2}\right), i=1,\ldots, N_t ,
\end{aligned}
\right.  \end{aligned}\label{eq:code-boook}
\end{equation}
\nid  for the case that $\phi^\star$ is in the $k$-th sub-interval.

Since the optimal $\phi^\star$ must be located in one of $N_t$ sub-intervals, we can obtain $N_t$ conditionally optimal  precoders by examining all $N_t$ sub-intervals and construct a candidate precoder set $\mathcal{F}_l$ as \be \mathcal{F}_l \triangleq \{\mathbf{f}_{l,1}, \ldots, \mathbf{f}_{l,{N_t}}\},  \ee \nid which must contain the optimal precoder $\mathbf{f}_{RF,l}^\star$. Therefore, without loss of performance, the problem in (\ref{eq:analog precoder design1}) can be transformed to an equivalent maximization task over only the set $\mathcal{F}_l$
\be
 \mathbf{f}_{{RF},{l}}^{\star} = \textrm{arg}  \underset{\mathbf{f}_{{RF},l}  \in  \mathcal{F}_l }{\textrm{max}}  \left| \mathbf{f}_{RF,l}^H \mathbf{g}_{l,1} \right| ,  \label{eq:analog precoder design3}
\ee
\nid which has linear complexity $\mathcal{O}(N_t)$.
Similarly, we can also construct a candidate analog combiner set $\mathcal{W}_l$  and obtain $\mathbf{w}_{{RF},{l}}^{\star}$ by the same procedure.

The rank-1 solution returned by (\ref{eq:analog precoder design3}) is based on the rank-1 approximation of the interference-included equivalent channel $\mathbf{Q}_l$. The approximation of $\mathbf{Q}_l$ may cause a performance degradation when we revisit the original problem (\ref{eq:beam sel_1-bit}).
Therefore, in order to enhance the performance, we propose to jointly select the precoder and combiner over candidate sets $\mathcal{F}_l$ and $\mathcal{W}_l$ as
\begin{eqnarray}
&\left\{\mathbf{f}_{{RF},{l}}^\star,\mathbf{w}_{{RF},{l}}^\star  \right\}=\textrm{arg}\underset{ \substack{\mathbf{f}_{{RF},l}  \in \mathcal{F}_l\\ \mathbf{w}_{{RF},l}  \in \mathcal{W}_l}}{\textrm{max}} \left| \mathbf{w}_{RF,l}^H \mathbf{Q}_l\mathbf{f}_{RF,l} \right| \label{eq:beam_sel_final}
\end{eqnarray}
which may return the rank-1 or a better solution with quadratic complexity $\mathcal{O}(N_t N_r)$. This low-complexity analog beamformer design with one-bit resolution PSs is summarized in Algorithm \ref{alg:one bit}.

\begin{algorithm}[htb!]
\caption {One-Bit Resolution Analog Beamformer Design}
\label{alg:one bit}
\begin{algorithmic}[1]
\REQUIRE  $\mathbf{Q}_l$.
\ENSURE   $\mathbf{f}_{{RF},{l}}^\star$  and $\mathbf{w}_{{RF},{l}}^\star$.
\STATE Calculate $\mathbf{p}_{l,1}$ and $\mathbf{g}_{l,1}$ by an SVD of $\mathbf{Q}_l$.
\STATE Define the angles $\widehat{\psi_i}$, $i=1, \ldots, N_t$, by (\ref{eq:psi_i}).
\STATE Map $\widehat{\psi_i}$ to $\widetilde{\psi_i}$, $i=1,\ldots, N_t$,  in an ascending
order.
\FOR {$k=1:N_t$}
\STATE Obtain $\mathbf{\widetilde{f}}_{l,k}$ by (\ref{eq:f_tilde}).
\STATE Obtain $\mathbf{\widehat{f}}_{l,k}$ from  $\mathbf{\widetilde{f}}_{l,k}$ based on inverse mapping from $\widetilde{\psi_i}$ to $\widehat{\psi_i}$, $i=1,\ldots, N_t$.
\STATE Obtain $\mathbf{f}_{l,k}$ from $\mathbf{\widehat{f}}_{l,k}$ by (\ref{eq:code-boook}).
\ENDFOR
\STATE Construct $\mathcal{F}_{l} = \{\mathbf{f}_{l,1}, \ldots, \mathbf{f}_{l,N_t} \}$.
\STATE Construct $\mathcal{W}_{l}$ by a similar procedure as Steps 2-9.
\STATE Find the optimal $\mathbf{f}_{{RF},{l}}^\star$ and $\mathbf{w}_{{RF},{l}}^\star$ by (\ref{eq:beam_sel_final}).
\end{algorithmic}
\end{algorithm}

\section{Hybrid Precoder and Combiner Design for Multiuser mmWave MIMO Systems}
\label{sec:Proposed Design Multiuser}

In this section, we consider a mmWave multiuser MIMO uplink system and extend the low-resolution hybrid precoder and combiner designs proposed in the previous sections to the multiuser system.

\subsection{System Model and Problem Formulation}

\begin{figure*}[!t]
\centering
\vspace{-0.0 cm}
\includegraphics[width= 5.6 in]{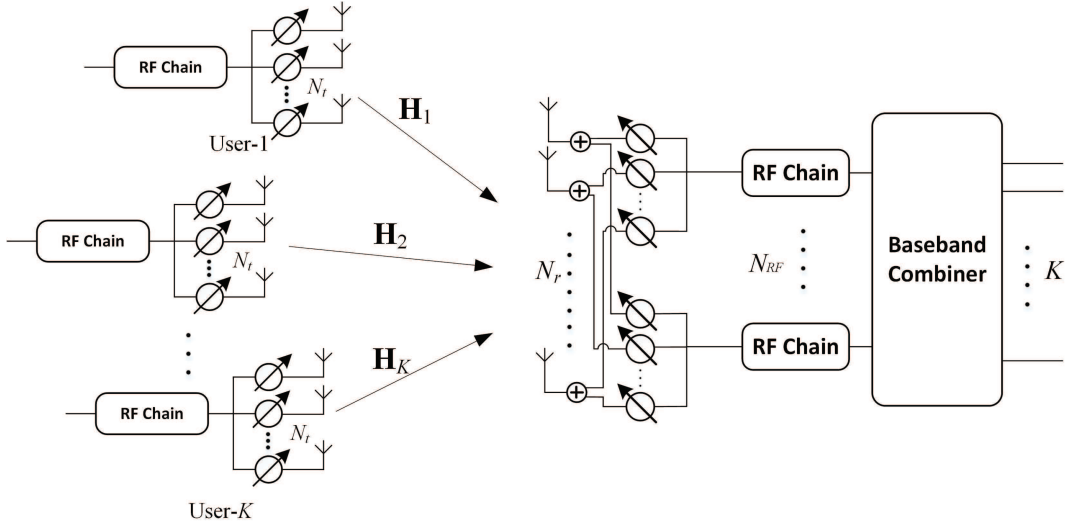}
\caption{The multiuser mmWave MIMO  system using hybrid precoder and combiner.}\label{fig:MU_system_model}\vspace{-0.0 cm}
\end{figure*}

We consider a multiuser mmWave MIMO uplink system as presented in Fig. \ref{fig:MU_system_model}, where a base-station (BS) is equipped with $N_r$ antennas and $N_{RF}$ RF chains and simultaneously serves $K$ mobile users. Due to power consumption and hardware limitations, each mobile user has $N_t$ antennas and a single RF chain to transmit only one data stream to the BS. We further assume the number of RF chains at the BS is equal to the number of users, i.e. $N_{RF} = K$.

Let $\mathbf{f}_{RF,k}$ be the analog precoder of the $k$-th user, where each element of $\mathbf{f}_{RF,k}$ has a constant magnitude $\frac{1}{\sqrt{N_t}}$ and low-resolution discrete phases, i.e. $\mathbf{f}_{RF,k}(i) \in \mathcal{F}$, $\forall i=1,\ldots, N_t$. The transmitted signal of the $k$-th user after precoding can be formulated as
\begin{equation}
\mathbf{x}_k=\sqrt{P_k}\mathbf{f}_{RF,k}s_k
\end{equation}
where $s_k$ is the symbol of the $k$-th user, $\mathbb{E}\{\left|s_k\right|^2\}=1$, and $P_k$ is the $k$-th user's transmit power.

Let $\mathbf{H}_k \in \mathbb{C}^{N_r\times N_t},~k=1,\ldots,K$, denote the uplink channel from the $k$-th user to the BS.
The received signal at the BS can be written as
\begin{equation}
\mathbf{r}=\sum_{k=1}^{K} \sqrt{P_k}\mathbf{H}_k\mathbf{f}_{RF,k}s_k+\mathbf{n}  \end{equation}
\nid where $\mathbf{n}\thicksim\mathcal{CN}(\mathbf{0},\sigma^2\mathbf{I}_{N_r})$ is complex Gaussian noise.
The BS first applies an $N_r\times K$ analog combining matrix $\mathbf{W}_{RF}\triangleq[\mathbf{w}_{{RF},1} \ldots \mathbf{w}_{{RF},K}]$ to process the received signal, in which the analog combiner $\mathbf{w}_{RF,k}$ corresponding to the $k$-th user is also implemented by low-resolution PSs, i.e. $\mathbf{w}_{{RF},k}(j) \in \mathcal{W}$, $j=1, \ldots, N_r$.
Then, a baseband digital combiner $ \mathbf{w}_{{BB},k} \in \mathbb{C}^{K \times 1}$ is employed to retrieve the information of the $k$-th user.  Let $\mathbf{w}_k\triangleq\mathbf{W}_{RF}\mathbf{w}_{{BB},k}$ denote the hybrid combiner corresponding to the $k$-th user. After the combining process at the BS, the estimated symbol of the $k$-th user can be expressed as
\begin{equation}
\hat{s}_k=\sqrt{P_k}\mathbf{w}^H_k\mathbf{H}_k\mathbf{f}_{RF,k}s_k+\mathbf{w}^H_k\sum_{\substack{i=1\\i\neq k}}^{K} \sqrt{P_i}\mathbf{H}_i\mathbf{f}_{RF,i}  s_i+\mathbf{w}^H_k\mathbf{n}.
\label{eq:received signal}
\end{equation}

Given the received signal at the BS in (\ref{eq:received signal}), the signal-to-interference-plus-noise ratio (SINR) of the $k$-th user can be expressed as
\begin{equation}
\begin{aligned}
&\gamma_k=\frac{\mid \sqrt{P_k}\mathbf{w}^H_k\mathbf{H}_k\mathbf{f}_{RF,k} \mid^2}{\sum\limits_{\substack{i=1,i\neq k}}^K\mid \sqrt{P_i}\mathbf{w}^H_k\mathbf{H}_i\mathbf{f}_{RF,i} \mid^2+\sigma^2\parallel\mathbf{w}_k\parallel^2}
\label{eq:sinr}
\end{aligned}
\end{equation}
and the achievable sum-rate of the multiuser uplink system is
\begin{equation}
R_{u}=\sum_{k=1}^{K}\log(1+\gamma_k).
\end{equation}

\nid We aim to jointly design the analog precoders and combiners implemented by low-resolution PSs as well as the digital combiners to maximize the sum-rate of the uplink multiuser system:
\begin{equation}
\begin{aligned}
&\hspace{-0.6 cm}\left\{\left\{\mathbf{w}_{{RF},k}^\star, \mathbf{w}_{{BB},k}^\star, \mathbf{f}_{RF,k}^\star \right\}_{k=1}^K\right\}
=\arg\max\sum_{k=1}^{K}
\log \left( 1+\gamma_k\right) \\
&\hspace{1.0 cm} \textrm{s. t.} ~~~~ \mathbf{f}_{RF,k}(i) \in \mathcal{F},~  \forall k, i,\\
&\hspace{2.0 cm} \mathbf{w}_{{RF},k}(j) \in \mathcal{W},~  \forall k, j.
\label{eq:objective function MU}
\end{aligned}
\end{equation}

\subsection{Low-Resolution Hybrid Precoder and Combiner Design}

Obviously, the optimization problem (\ref{eq:objective function MU}) cannot be directly solved. Thus, we adopt an approach similar to \cite{Wang 17} and propose to successively design the low-resolution analog beamformer pair for each user, aiming at enhancing the channel gain as well as suppressing the inter-user interference. Then, the baseband combiner at the BS is calculated to further mitigate the interference and maximize the sum-rate.

In particular,  for the first user, the analog precoder and combiner pair is designed to maximize the corresponding channel gain, which can be formulated as follows:
\begin{equation}
\begin{aligned}
&\hspace{0.4 cm}\left\{ \mathbf{w}^\star_{{RF},1}, \mathbf{f}^\star_{{RF},1} \right\} = \textrm{arg} \;
{\textrm{max}} \left| \mathbf{w}^H_{{RF},1}\mathbf{{H}}_1 \mathbf{f}_{{RF},1}\right|\\
&\hspace{1.7 cm}\textrm{s. t.} ~~~ \mathbf{f}_{{RF},1}(i) \in \mathcal{F},~  i=1,\ldots,N_t,\\
&\hspace{2.6 cm} \mathbf{w}_{{RF},1}(j) \in \mathcal{W},~  j=1,\ldots,N_r.\label{eq:beam design MU}
\end{aligned}
\end{equation}
This analog precoder and combiner design  problem can be efficiently solved by the algorithm presented in Sec. \ref{sec:low resolution analog beamformer} when low-resolution PSs are utilized, or the algorithm proposed in Sec. \ref{sec:low complexity scheme} if only one-bit resolution PSs are available. Then, the analog precoders $\mathbf{f}_{RF,k}$ and combiners $\mathbf{w}_{{RF},k}$, $k=2,3,\ldots,K $, for the remaining $K-1$ users are successively designed by an iterative procedure. In each iteration, we attempt to find the analog beamformer pair that suppresses the interference from the users whose analog beamformers have already been determined.
To achieve this goal, the channel of the user whose combiner is to be calculated is projected onto the space orthogonal to the collection of previously designed analog combiners. This approach leads to orthogonal analog combiners that suppress the inter-user interference.

Specifically, to design the $k$-th user's analog beamformer pair, we first extract the orthonormal components $\mathbf{d}_i$ of the previously determined analog combiners $\mathbf{w}^\star_{{RF},i}$, $i=1,\ldots,k-1$ by the Gram-Schmidt procedure:
\begin{equation}
\mathbf{q}_i=\mathbf{w}^\star_{{RF},i} -\sum\limits_{j=1}^{i-1}\mathbf{d}^H_j \mathbf{w}^\star_{{RF},i}\mathbf{d}_j , \end{equation}
\begin{equation}
  \mathbf{d}_i= \mathbf{q}_i/\|\mathbf{q}_i\|.
\end{equation}
Note that $\mathbf{d}_1 = \mathbf{w}^\star_{{RF},1}$ and $\mathbf{w}^\star_{{RF},1}$ is the analog combiner calculated for the first user. 
Then, the combiner components are removed from the $k$-th user's  channel to obtain the modified channel $\mathbf{\widehat{H}}_k$ as
\begin{equation}
\mathbf{\widehat{H}}_k=\left( \mathbf{I}_{N_r}-\sum_{i=1}^{k-1}\mathbf{d}_i \mathbf{d}_i^H \right)\mathbf{H}_k.
\end{equation}
Finally, based on the modified channel $\mathbf{\widehat{H}}_k$, the analog beamformer pair for the $k$-th user is found by solving the following optimization using the algorithms proposed in the previous sections:
\begin{equation}
\begin{aligned}
&\hspace{0.4 cm}\left\{ \mathbf{w}^\star_{{RF},k}, \mathbf{f}^\star_{{RF},k} \right\} = \textrm{arg} \;
{\textrm{max}} \left| \mathbf{w}^H_{{RF},k}\mathbf{\widehat{H}}_k \mathbf{f}_{RF,k}\right|\\
&\hspace{1.7 cm}\textrm{s. t.} ~~~ \mathbf{f}_{RF,k}(i) \in \mathcal{F},~  i=1,\ldots,N_t,\\
&\hspace{2.6 cm} \mathbf{w}_{{RF},k}(j) \in \mathcal{W},~  j=1,\ldots,N_r.\label{eq:beam design MU1}
\end{aligned}
\end{equation}

After finding the analog beamformers for all users, the effective baseband channel for each user can be obtained as  $\mathbf{h}^{e}_k\triangleq\sqrt{P_k}\left(\mathbf{W}^\star_{RF}\right)^H \mathbf{H}_k\mathbf{f}^\star_{{RF},k}$. Then, a minimum mean square error (MMSE) baseband digital combiner for the $k$-th user is employed to further suppress the interference: 
\begin{equation}
\begin{aligned}
 \mathbf{w}^\star_{{BB},k}= \left[\mathbf{{H}}^e(\mathbf{{H}}^e)^H+\sigma^2 \left(\mathbf{W}^\star_{RF}\right)^H\mathbf{W}^\star_{RF}\right]^{-1} \mathbf{h}^{e}_k,\label{eq:digital combiner}
\end{aligned}
\end{equation}
where $\mathbf{H}^{e}\triangleq[\mathbf{h}^{e}_1,\ldots,\mathbf{h}^{e}_K]$. The proposed low-resolution hybrid precoder and combiner design for multiuser mmWave systems is summarized in Algorithm \ref{alg:MU}.

\begin{algorithm}[htb!]
\caption{Low-Resolution Hybrid Precoder and Combiner Design for Multiuser mmWave Systems}
\label{alg:MU}
\begin{algorithmic}[1]
\REQUIRE $\mathcal{F}$, $\mathcal{W}$, $\mathbf{H}_k$, $k=1,\ldots,K$.
\ENSURE $\mathbf{f}_{RF,k}^\star$, $\mathbf{w}^\star_{{RF},k}$, $\mathbf{w}^\star_{BB,k}$, $k=1,\ldots,K$.
\STATE Obtain $\mathbf{w}^\star_{{RF},1}$ and $\mathbf{f}^\star_{{RF},1}$  for user-1 by solving
\begin{equation}
\hspace{-1 cm}  \left\{ \mathbf{w}^\star_{{RF},1}, \mathbf{f}_{RF,1}^\star \right\} = \textrm{arg} \underset{\substack{\mathbf{w}_{{RF},1}(i) \in \mathcal{W} \\ \mathbf{f}_{RF,1}(j) \in \mathcal{F}}} {\textrm{max}} | \mathbf{w}^H_{{RF},1}\mathbf{H}_1\mathbf{f}_{RF,1} | . \nonumber
\end{equation}
\STATE $\mathbf{d}_1=\mathbf{w}^\star_{{RF},1}$.
\FOR {$k=2:K$}
\STATE $\mathbf{\widehat{H}}_k=\left( \mathbf{I}_{N_r}-\sum\limits_{i=1}^{k-1}\mathbf{d}_i \mathbf{d}_i^H \right)\mathbf{H}_k$.
\STATE Obtain $\mathbf{w}^\star_{{RF},k}$ and $\mathbf{f}^\star_{{RF},k}$ for user-$k$ by solving
\begin{equation}
\hspace{-0.5 cm} \left\{ \mathbf{w}^\star_{{RF},k}, \mathbf{f}^\star_{{RF},k} \right\} = \textrm{arg} \underset{\substack{\mathbf{w}_{{RF},k}(i) \in \mathcal{W} \\ \mathbf{f}_{RF,k}(j) \in \mathcal{F}}} {\textrm{max}} | \mathbf{w}^H_{{RF},k}\mathbf{\widehat{H}}_k\mathbf{f}_{RF,k}|. \nonumber
\end{equation}
\STATE $\mathbf{q}_k=\mathbf{w}^\star_{{RF},k} -\sum\limits_{i=1}^{k-1}\mathbf{d}^H_i \mathbf{w}^\star_{{RF},k}\mathbf{d}_i$;
\STATE $\mathbf{d}_k= \mathbf{q}_k/\|\mathbf{q}_k\|$.
\ENDFOR
\STATE Obtain digital combiners $\mathbf{w}^\star_{{BB},k}$, $k=1,\ldots, K$, by
\begin{equation}
\mathbf{w}^\star_{{BB},k}=\left[\mathbf{H}^e(\mathbf{H}^e)^H+
\sigma^2\left(\mathbf{W}^\star_{RF}\right)^H\mathbf{W}^\star_{RF}\right]^{-1}\mathbf{h}^e_k.  \nonumber
\end{equation}
\end{algorithmic}
\end{algorithm}

\section{Simulation Results}
\label{sc:Simulation}

\begin{figure}[!t]
\centering
  \includegraphics[width=3.5 in]{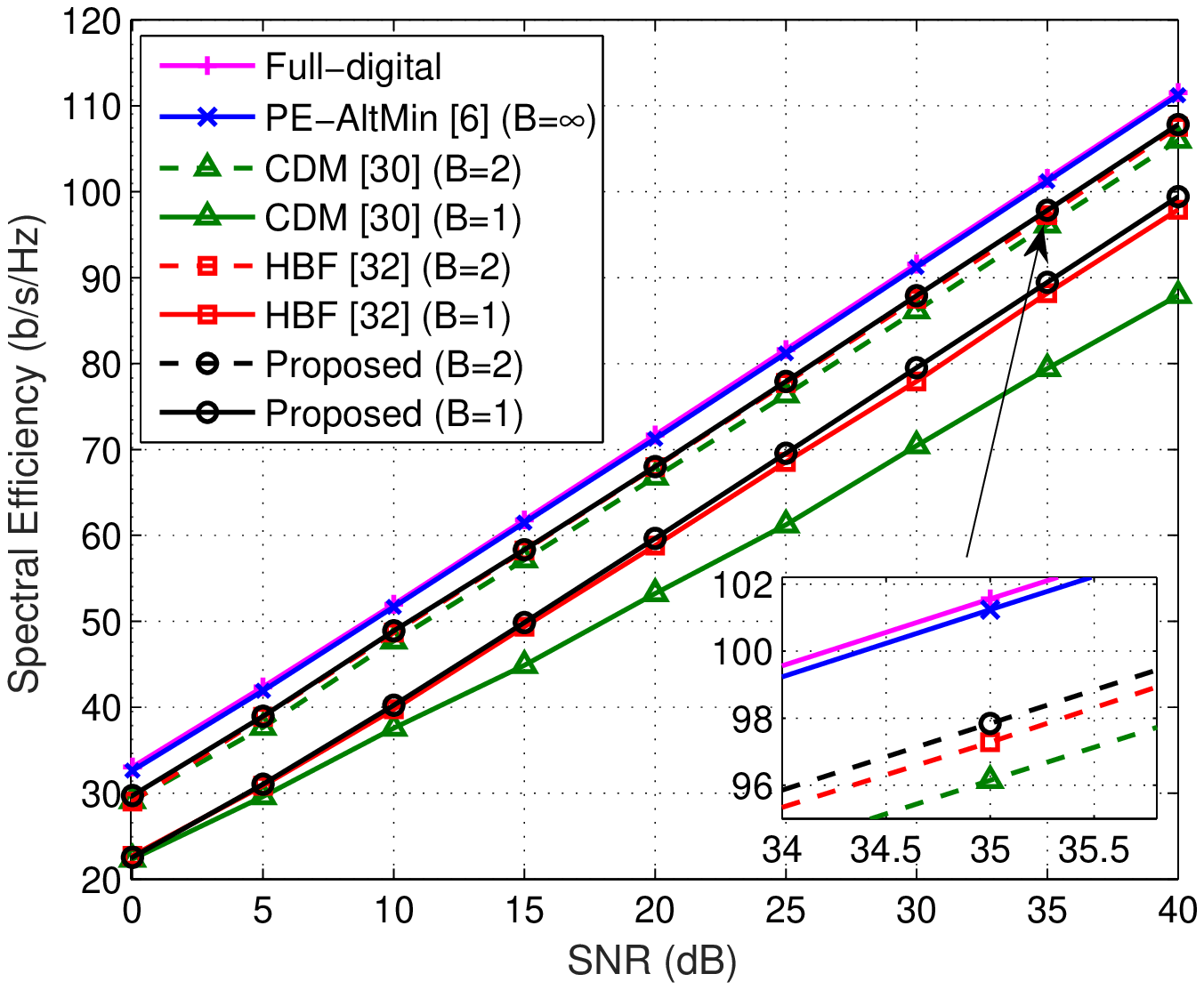}
  \vspace{-0.4 cm}
  \caption{Spectral efficiency versus SNR ($N_t=64$, $N_r=64$, $N_t^{RF}= N_r^{RF}=6$, $N_s=6$).}\label{fig:su_c_vs_snr}
  \vspace{0.3 cm}
  \includegraphics[width=3.5 in]{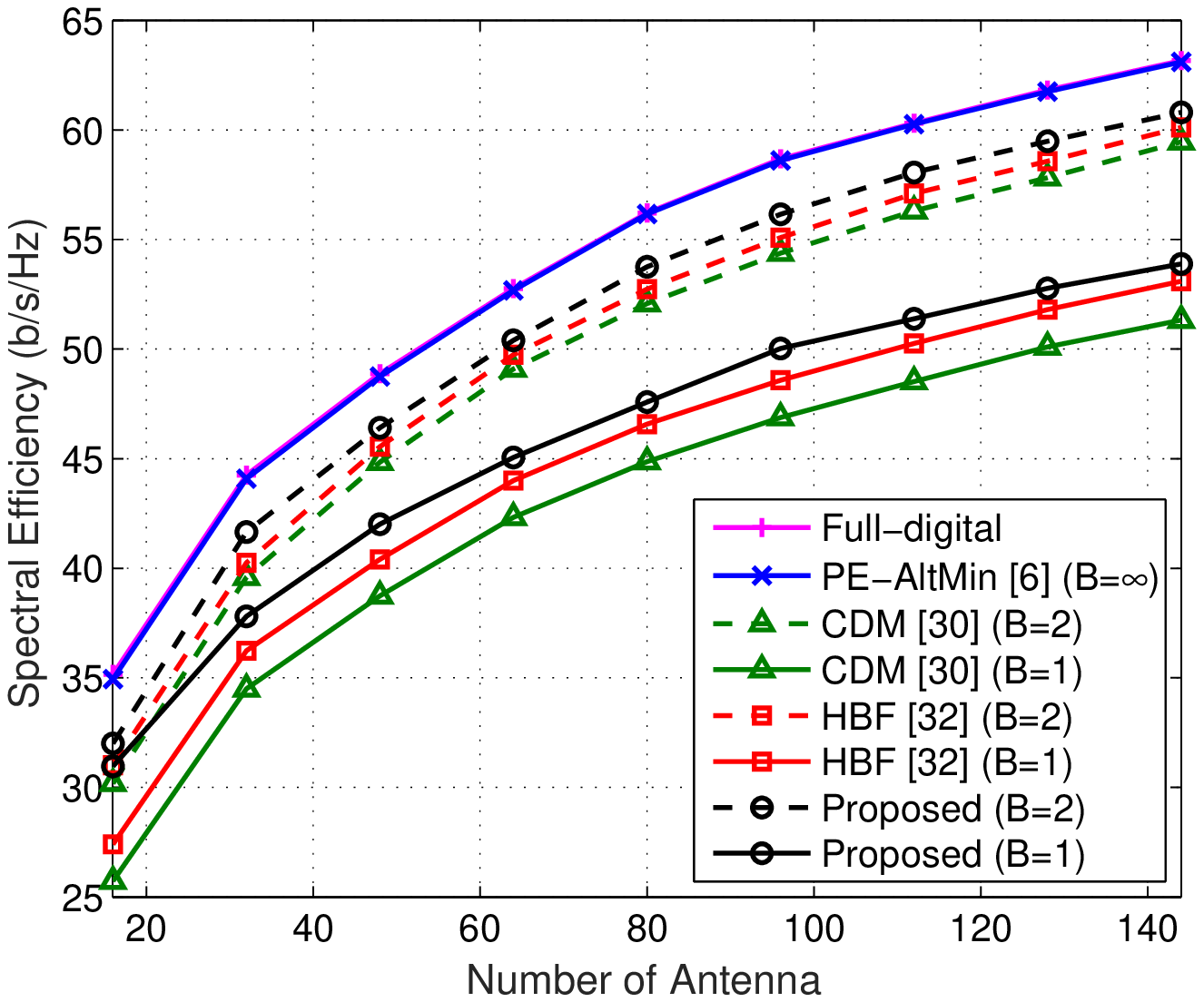}
  \vspace{-0.4 cm}
  \caption{Spectral efficiency versus number of antenna ($N_t^{RF}= N_r^{RF}=4$, $N_s=4$, SNR = $20$dB).}\label{fig:su_c_vs_n}\vspace{-0.0 cm}
\end{figure}

\begin{figure}[!t]
\centering
  \includegraphics[width=3.5 in]{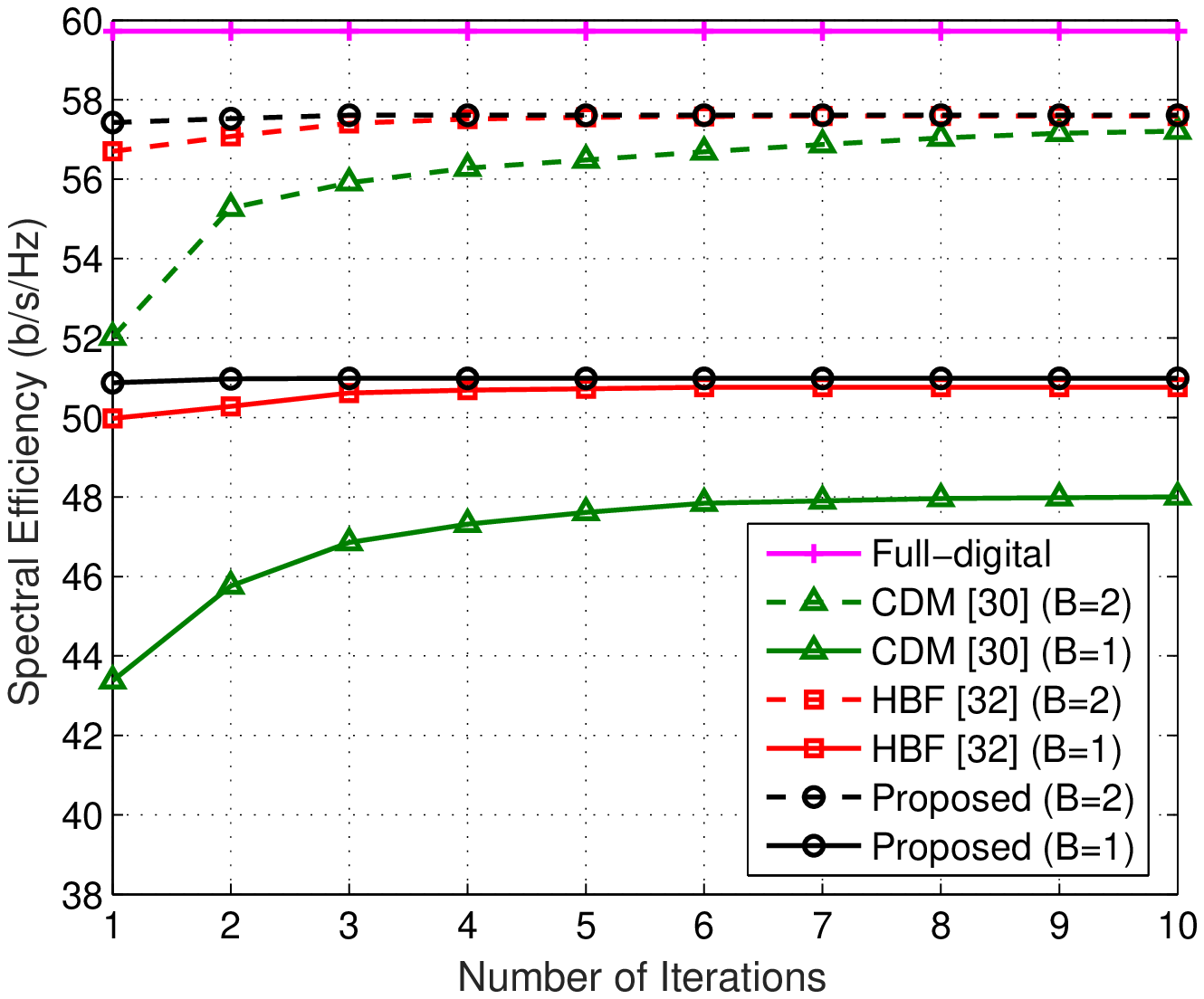}
  \vspace{-0.4 cm}
  \caption{Spectral efficiency versus number of iteration ($N_t=64, N_r=64$, $N_t^{RF}= N_r^{RF}=4$, $N_s=4$, SNR = $20$dB).}\label{fig:su_c_vs_niter}
  \vspace{0.3 cm}
  \includegraphics[width=3.5 in]{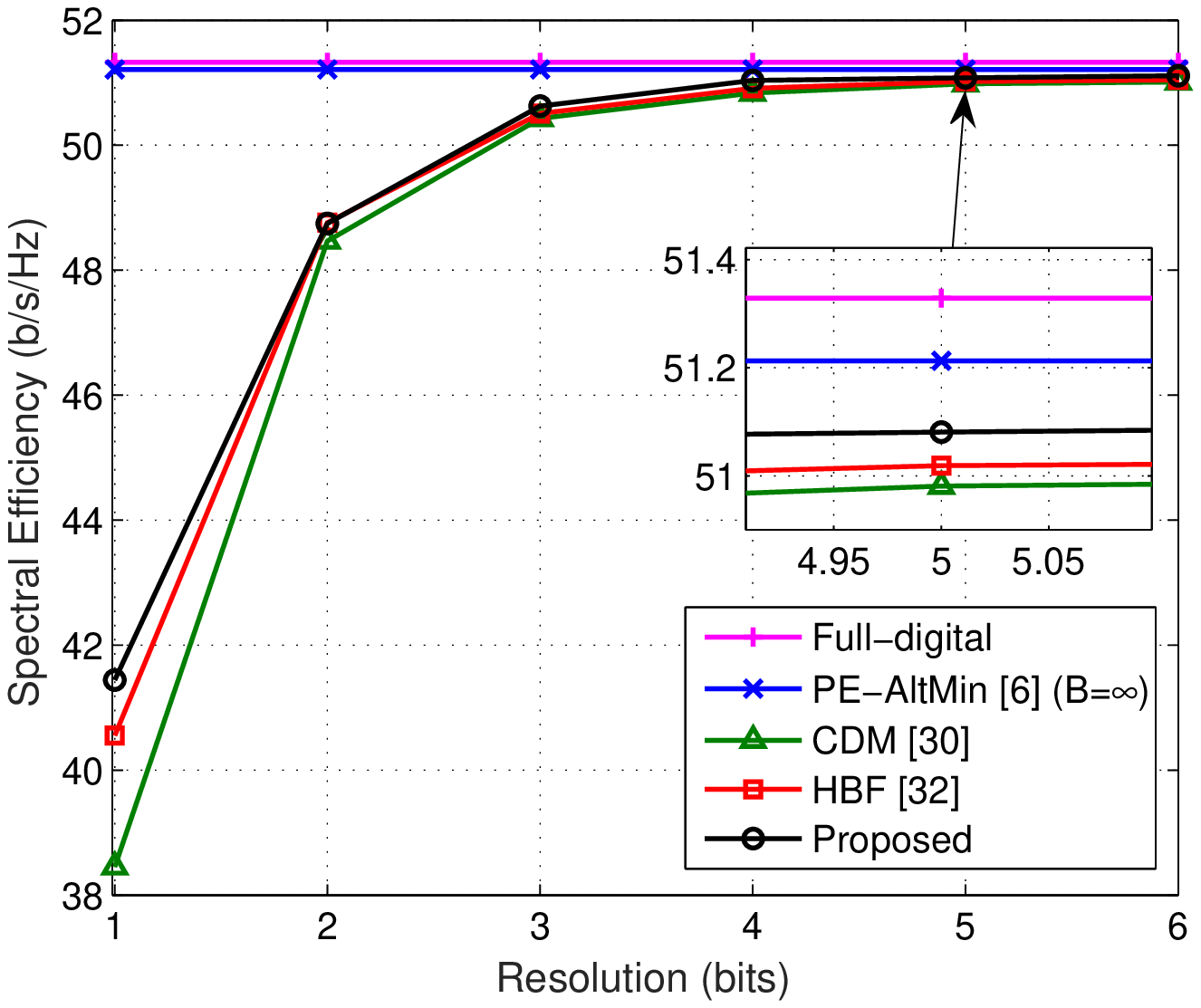}
   \vspace{-0.4 cm}
\caption{Spectral efficiency versus resolution of PSs ($N_t=64$, $N_r=64$, $N_t^{RF}= N_r^{RF}=4$, $N_s=4$, $\mathrm{SNR}=20$dB).}\label{fig:SU_C_vs_B}
\vspace{-0.0 cm}
\end{figure}

\begin{figure}[!t]
\centering
   \includegraphics[width=3.5 in]{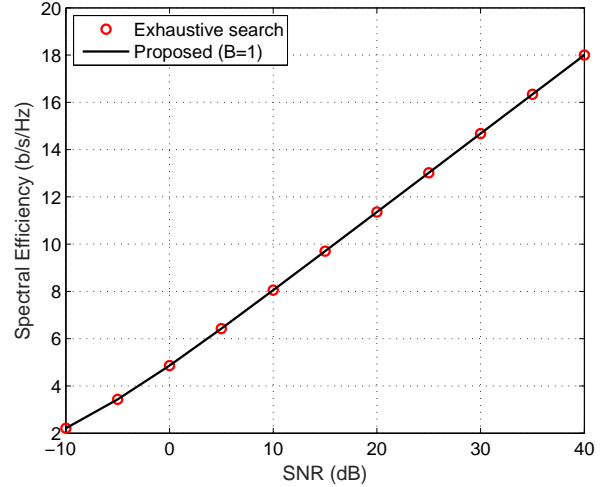}
  \vspace{-0.4 cm}
  \caption{Spectral efficiency versus SNR ($N_t=8$, $N_r=8$, $N_s=1$, $B=1$).}\label{fig:SU_ES_M8N8Ns1}

\end{figure}

In this section, we provide simulation results for the proposed joint hybrid precoder and combiner designs with low-resolution PSs for point-to-point mmWave systems as well as multiuser mmWave systems.
MmWave channels are expected to be sparse and have a limited number of propagation paths. In the simulations, we adopt a geometric channel model with $L$ paths \cite{Sohrabi 16}.
In particular, the discrete-time narrow-band mmWave channel $\mathbf{H}$ is formulated as
\begin{equation}
\mathbf{H}=\sqrt{\frac{N_t N_r}{L}} \sum_{i=1}^{L} \alpha_{i}\mathbf{a}_{r}(\theta_{i}^{r})\mathbf{a}_{t}(\theta_{i}^{t})^H
\label{eq:channel_model}
\end{equation}
where $\alpha_{i}\thicksim \mathcal{CN}(0,\frac{1}{L})$ are the independent and identically distributed complex gains of the $i$-th propagation path (ray)
$\theta_{i}^{t}$ and $\theta_{i}^{r}$ $\in [-\frac{\pi}{2}, \frac{\pi}{2}]$
are the angles of departure (AoD) and the angles of arrival (AoA), respectively.
Finally, the array response vectors $\mathbf{a}_{t}(\theta^t)$ and $\mathbf{a}_{r}(\theta^r)$ depend on the antenna array geometry. We assume that the commonly used uniform linear arrays (ULAs) are employed, and the transmit antenna array response vector $\mathbf{a}_t(\theta^t)$ and the receive antenna array response vector $\mathbf{a}_r(\theta^r)$ can be written as
\begin{equation}
\hspace{-0.0 cm}\mathbf{a}_t(\theta^t)=\frac{1}{\sqrt{N_t}}[1, {e}^{j\frac{2\pi}{\lambda}d\sin(\theta^t)}, \ldots , {e}^{j(N_t-1)\frac{2\pi}{\lambda}d\sin(\theta^t)}]^T , 
\end{equation}
\begin{equation}
\hspace{-0.0 cm}\mathbf{a}_r(\theta^r)=\frac{1}{\sqrt{N_r}}[1, {e}^{j\frac{2\pi}{\lambda}d\sin(\theta^r)}, \ldots , {e}^{j(N_r-1)\frac{2\pi}{\lambda}d\sin(\theta^r)}]^T , 
\end{equation}
\nid respectively, where $\lambda$ is the signal wavelength, and $d$ is the distance between antenna elements. In the following simulations, we consider an environment with $L = 6$ scatterers between the transmitter and the receiver. The antenna spacing is $d=\frac{\lambda}{2}$.

%

\subsection{Simulation Results of a Point-to-Point mmWave System}

We first consider a point-to-point mmWave communication system, in which the transmitter and receiver are both equipped with $64$-antenna ULAs.  The number of RF chains at the transmitter and receiver are $N_{t}^{RF} = N_{r}^{RF} = 6$, so the number of data streams is also assumed to be $N_s = 6$.

Fig. \ref{fig:su_c_vs_snr} shows the average spectral efficiency versus SNR over $10^6$ channel realizations.
We evaluate the spectral efficiency of the algorithm proposed in Sec. III for the case of 2-bit ($B=2$) resolution PSs and the algorithm proposed in Sec. IV for the case of 1-bit ($B=1$) resolution PSs.
For comparison purposes, we also plot the spectral efficiency of two state-of-the-art low-resolution hybrid beamformer designs: the coordinate descent method (CDM) algorithm in \cite{Chen TVT 17} and the hybrid beamforming (HBF) algorithm in \cite{Sohrabi 16}.
To the best of our knowledge, the algorithm in \cite{Sohrabi 16} achieves the best performance with low-resolution PSs in the existing literature.
The performance of a fully digital approach using SVD-based beamforming and the hybrid beamforming scheme with infinite-resolution ($B=\infty$) PSs using the phase extraction (PE-AltMin) algorithm in \cite{Yu JSAC 16} are also included as performance benchmarks.
Fig. \ref{fig:su_c_vs_snr} illustrates that the proposed algorithm outperforms the competitors, particularly for the case of 1-bit resolution PSs.
Moreover, it can be observed that the proposed algorithm with $B = 2$ achieves performance close to optimal full-digital beamforming and hybrid beamforming with infinite-resolution PSs.
For additional simulation validation, Fig. \ref{fig:su_c_vs_n} illustrates the spectral efficiency versus the number of antennas and similar conclusions can be drawn.

In order to illustrate the convergence of the proposed algorithm, we show the spectral efficiency versus the number of iterations in Fig. \ref{fig:su_c_vs_niter}, which also includes other algorithms for comparison. It is observed that our proposed algorithms converge faster than the other two iterative schemes, which is a highly favorable property.
In Fig. \ref{fig:SU_C_vs_B}, we show the spectral efficiency as a function of $B$ to illustrate the impact of the resolution of PSs on the spectral efficiency.
As expected, increasing the PS resolution will improve the system performance, but using only $B=3$ bits is sufficient to closely approach the performance of the ideal unquantized case. Beyond $B=3$, the additional cost and complexity associated with using higher-resolution PSs is not warranted given the very marginal increase in spectral efficiency. Moreover, our proposed algorithms outperform the other two low-resolution beamforming methods for all PS resolutions.

\begin{figure}[!t]
\centering
\includegraphics[width=3.5 in]{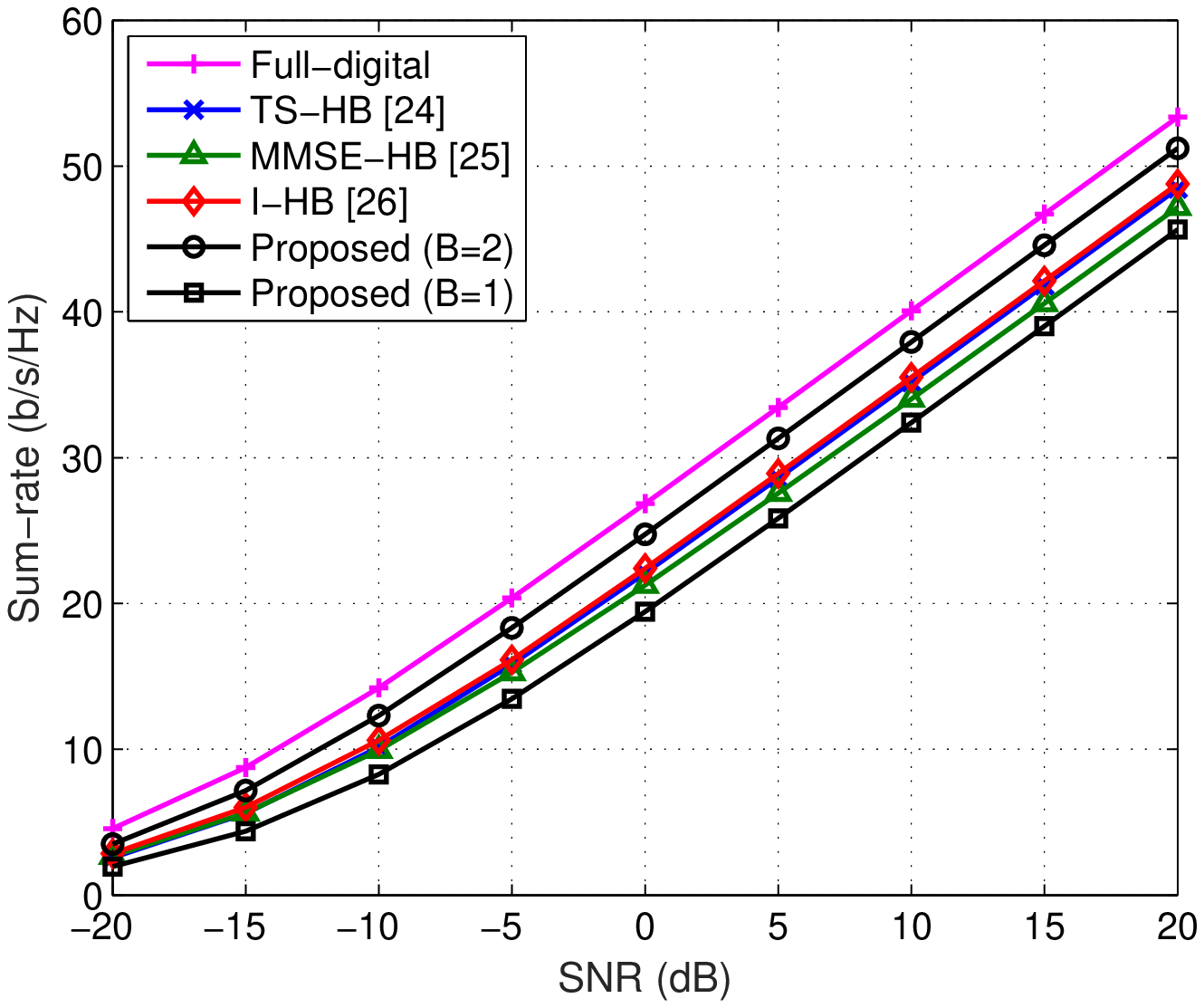}
  \vspace{-0.4 cm}
  \caption{Spectral efficiency versus SNR ($N_t=64$, $N_r=16$, $N_t^{RF}=4$, $K=4$).}\label{fig:MU_M64N16K4}
  \vspace{0.3 cm}
\includegraphics[width=3.5 in]{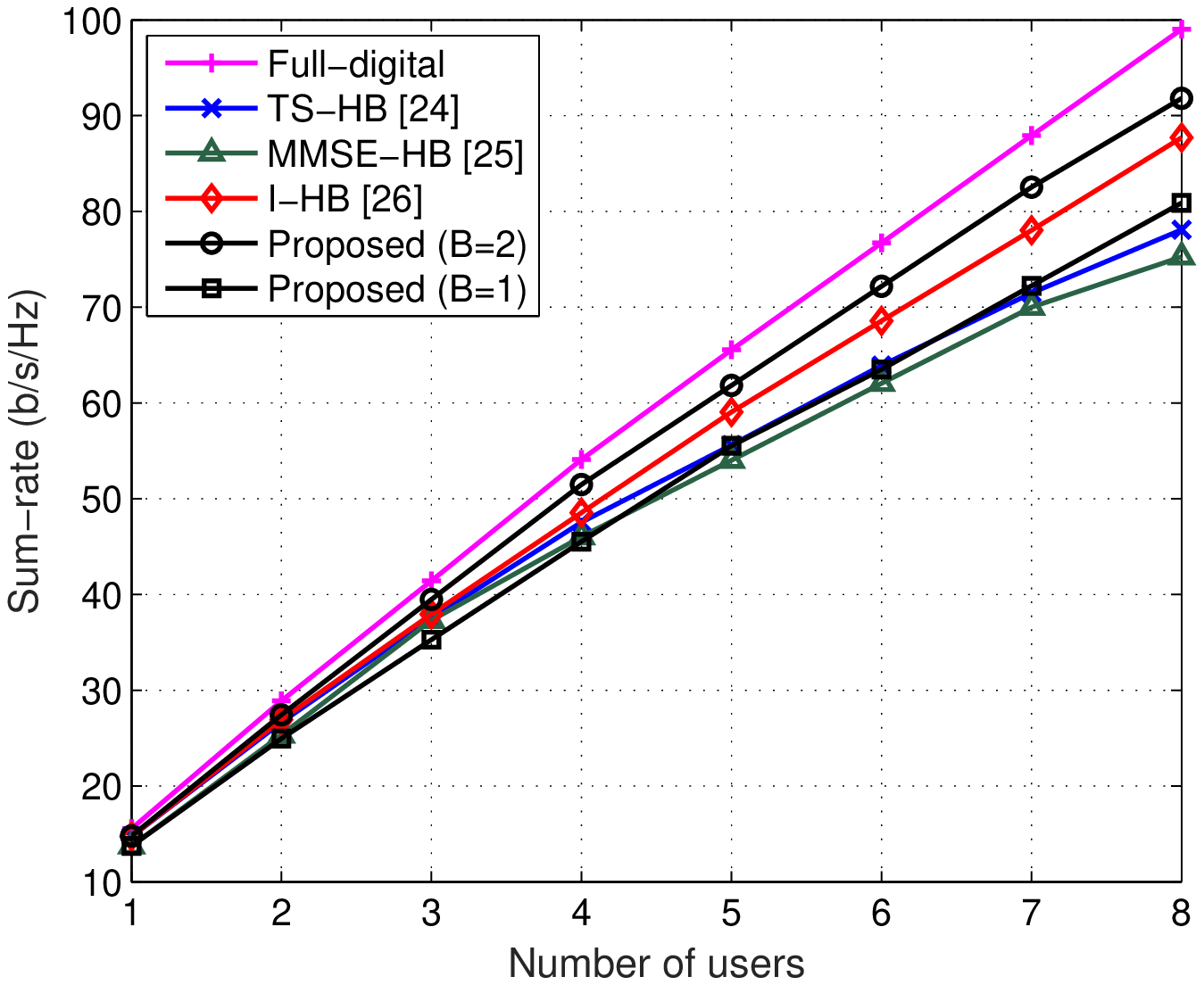}
  \vspace{-0.4 cm}
  \caption{Spectral efficiency versus $K$ ($N_t=64, N_r=16$, $N_t^{RF}=K$, SNR = $20$dB).}\label{fig:MU_R_vs_Ns_SNR20}\vspace{-0.0 cm}
\end{figure}

To examine the impact of the approximations used in deriving the proposed one-bit resolution hybrid beamformer scheme, in Fig. \ref{fig:SU_ES_M8N8Ns1} we compare it with the optimal exhaustive search approach.
The number of  antennas at both transmitter and receiver is chosen to be 8 and the number of data streams is $N_s=1$. A relatively simple case is examined here due to the exponential complexity of the exhaustive search method.
We see from Fig. \ref{fig:SU_ES_M8N8Ns1} that the spectral efficiency achieved by the proposed algorithm is the same as that of the optimal exhaustive search method, suggesting that the proposed hybrid beamforming algorithm with one-bit resolution PSs can provide optimal or near-optimal performance.

\subsection{Simulation Results of a Multiuser mmWave System}

Next, we evaluate the performance of the proposed low-resolution beamformer algorithm in a multiuser uplink system. We assume there are $K=4$ users, each of which is equipped with $N_t=16$ antennas and only one RF chain to transmit a single data stream. The BS has $N_r=64$ antennas and $N_r^{RF}=4$ RF chains.
Fig. \ref{fig:MU_M64N16K4} illustrates the sum-rate versus SNR for various hybrid beamformer designs. In particular, we include three state-of-the-art multiuser hybrid beamforming approaches for comparison: \textit{i}) two-stage hybrid beamforming (TS-HB) in \cite{Alkhateeb 15}, \textit{ii}) MMSE-based hybrid beamforming (MMSE-HB) in \cite{Nguyen ICC 16}, and \textit{iii}) iterative hybrid beamforming (I-HB) in \cite{Wang 17}.
All three algorithms are codebook-based approaches and the size of the beamsteering codebook is set at $32$ (i.e. $B=5$ quantization bits) for fairness of the comparison.
It can be observed from Fig. \ref{fig:MU_M64N16K4} that our proposed low-resolution hybrid beamforming design outperforms the other three algorithms using only 2-bit resolution PSs. Moreover, the performance with 1-bit resolution PSs is also comparable.
Fig. \ref{fig:MU_R_vs_Ns_SNR20} further shows the sum-rate versus the number of users $K$.
From Fig. \ref{fig:MU_R_vs_Ns_SNR20}, we see that our proposed algorithm with 2-bit resolution PSs always outperforms the other codebook-based  algorithms.
Furthermore, even with 1-bit resolution PSs, the proposed algorithm can still achieve competitive performance compared with the TS-HB and MMSE-HB approaches when $K>5$.

\section{Conclusions}
\label{sc:Conclusions}

This paper considered the problem of hybrid precoder and combiner design for mmWave MIMO systems with low-resolution quantized PSs.
We proposed an efficient iterative algorithm which  successively designs the low-resolution analog precoder and combiner pair for each
data stream.
Then, the digital precoder and combiner were computed based on the obtained effective baseband channel to further enhance the spectral efficiency.
The design of low-resolution hybrid beamformers for multiuser MIMO communication systems was also investigated.
Simulation results verified the effectiveness of
the proposed algorithms, particularly for scenarios in which one-bit resolution phase shifters are used.

%
%

\begin{appendices}

\section{Proof of Proposition 1}

The optimization problem (\ref{eq:beam sel}) can be equivalently formulated as
\begin{equation}
{\textrm{max}}\left| \frac{1}{\sqrt{N_tN_r}} \sum_{j=1}^{N_r}e^{j\varphi_j^l}\sum_{i=1}^{N_t} e^{j\vartheta_i^l}\mathbf{Q}_l(j,i) \right|.
\label{eq:phase match}
\end{equation}
By discarding the constant coefficient $1/\sqrt{N_tN_r}$, (\ref{eq:phase match}) can be further transformed as
\begin{equation}
{\textrm{max}}\left| \sum_{j=1}^{N_r}e^{j\varphi_j^l} \left\{e^{j\vartheta_i^l}\mathbf{Q}_l(j,i)+ \sum_{u\neq i}^{N_t} e^{j\vartheta_u^l}\mathbf{Q}_l(j,u)\right\} \right|.
\label{eq:phase match1}
\end{equation}
Since the term $e^{j \vartheta_i^l}$ does not involve the summation index $j$, it can be put outside the first summation, resulting in
\begin{equation}
{\textrm{max}}\left| e^{j\vartheta_i^l} \sum_{j=1}^{N_r}e^{j\varphi_j^l} \mathbf{Q}_l(j,i)+ \sum_{j=1}^{N_r}e^{j\varphi_j^l}\sum_{u\neq i}^{N_t} e^{j\vartheta_u^l}\mathbf{Q}_l(j,u) \right|.
\label{eq:phase match2}
\end{equation}
Obviously, the optimal value of $\vartheta_i^l$  makes the phases of the first and second term equal to obtain the largest amplitude, and (\ref{eq:opt phase f}) is proved. $\hfill$ $\blacksquare$

\end{appendices}

\end{document}